\def\year{2022}\relax
\documentclass[letterpaper]{article} 
\usepackage{aaai21}  
\usepackage{times}  
\usepackage{helvet} 
\usepackage{courier}  
\usepackage[hyphens]{url}  
\usepackage{graphicx} 
\def\UrlFont{\rm}  
\usepackage{natbib}  
\usepackage{caption} 
\usepackage{dcolumn}
\usepackage{multirow}
\usepackage{tabularx}
\usepackage{subcaption}
\usepackage{svg}
\usepackage{amsmath}
\newcolumntype{d}[1]{D{.}{.}{#1}}

\usepackage{svg}
\usepackage{balance}       
\usepackage{graphics}      
\usepackage[T1]{fontenc}   
\usepackage{txfonts}
\usepackage{mathptmx}
\usepackage{color}
\usepackage{booktabs}
\usepackage{textcomp}
\usepackage{float}

\usepackage{microtype}        
\usepackage{ccicons}          
\usepackage[utf8]{inputenc} 

\def\plaintitle{The Geography of Facebook Groups in the United States\footnote{To be presented at ICWSM 2023.}}
\def\plainauthor{Amaç Herdağdelen, Lada Adamic, Bogdan State}
\def\plainkeywords{social capital; social media; Facebook Groups}

\makeatletter
\def\url@leostyle{%
  \@ifundefined{selectfont}{
    \def\UrlFont{\sf}
  }{
    \def\UrlFont{\small\bf\ttfamily}
  }}
\makeatother
\def\pprw{8.5in}
\def\pprh{11in}

\definecolor{linkColor}{RGB}{6,125,233}
\usepackage{hyperref}
\hypersetup{
  pdftitle={\plaintitle},
  pdfauthor={\plainauthor},
  pdfkeywords={\plainkeywords},
  pdfdisplaydoctitle=true, 
  bookmarksnumbered,
  pdfstartview={FitH},
  colorlinks,
  citecolor=black,
  filecolor=black,
  linkcolor=black,
  urlcolor=linkColor,
  breaklinks=true,
  hypertexnames=false
}

\title{\plaintitle}
\author{Amaç Herdağdelen,\textsuperscript{1}
Lada Adamic,\textsuperscript{1}
Bogdan State\textsuperscript{2}\\
\textsuperscript{1}{Core Data Science, Meta}\\
\textsuperscript{2}{scie.nz}\\
amac@herdagdelen.com,
ladamic@gmail.com, 
bogdan@scie.nz}

\begin{document}
\maketitle

\begin{abstract}
We present a de-identified and aggregated dataset based on geographical patterns of Facebook Groups usage and demonstrate its association with measures of social capital. The dataset is aggregated at United States county level. Established spatial measures of social capital are known to vary across US counties. Their availability and recency depends on running costly surveys.  We examine to what extent a dataset based on usage patterns of Facebook Groups, which can be generated at regular intervals, could be used as a partial proxy by capturing local online associations. 

We identify four main latent factors that distinguish Facebook group engagement by county, obtained by exploratory factor analysis. The first captures small and private groups, dense with friendship connections. The second captures very local and small groups. The third captures non-local, large, public groups, with more age mixing. The fourth captures partially local groups of medium to large size. 
Only two of these factors, the first and third, correlate with offline community level social capital measures, while the second and fourth do not. Together and individually, the factors are predictive of offline social capital measures, even controlling for various demographic attributes of the counties. 

To our knowledge this is the first systematic test of the association between offline regional social capital and patterns of online community engagement in the same regions. By making the dataset available to the research community, we hope to contribute to the ongoing studies in social capital.
\end{abstract}

Understanding the norms and bonds that allow communities to act together constitutes one of the key focus areas for modern social science. The concept of \textit{social capital} has long been used to refer to a collection of measures at the individual, inter-personal and community level. These measures account for the strength of a diffuse fabric of trust relations, group affiliations and organizational structures. Social capital allows individuals to draw benefit from their connections, while also enabling communities to act collectively for common benefit~\cite{scott2005bowling}. The degree to which social capital varies between countries and communities has elicited a great deal of interest. In the United States in particular, the extent to which social capital measures vary between counties is a well-established sociological finding.
 
The rise of Internet-based social networking has facilitated the development of place-based virtual communities. Multiple platforms -- community forums, local subreddits, local Facebook pages, groups and events, Neighborly pages, to name but a few -- have emerged that cater to the need of virtual connection in local communities. Their importance has been bolstered by difficulty of in-person interaction during the COVID-19 pandemic. The development of such virtual arenas raises an interesting question as to the extent to which known differences in offline social capital are reproduced in the online world. 

Facebook -- the world's largest social network at the time of writing and the focus of our analysis -- is an important platform facilitating connections within local communities, in particular through the medium of the ``Facebook Groups'' product.  Facebook groups support a wide variety of local communities, for people bound together by shared neighborhood issues, hobbies, interests, or affiliations. These include groups corresponding to local organizations, such as neighborhood associations, scout troops, and sports clubs, to more informal associations such as those between parents of a cohort within a school district or a local meetup group for various hobbies and interests. Still other local groups have a commercial focus, such as ``for sale'' groups, while others support gift exchange as part of the ``buy nothing'' movement. 

It is not surprising to note that many Facebook groups' existence, membership, and interactions are shaped by offline contexts. For example, the number, size, and characteristics of Facebook groups may in part depend on the presence of offline organizations. Furthermore, interaction norms in the offline context may also be present online. This observation brings forth the expectation of community-level differences in social capital also being present in the online realm of local Facebook groups in particular, given the strong influence that local contexts are expected to have on these virtual entities. 

The question of how to operationalize and measure social capital remains an open one. Measures of social capital include \textit{generalized trust} (defined as the extent to which individuals believe unknown alters can be trusted; see \citeauthor{bjornskov2007determinants}, \citeyear{bjornskov2007determinants}), as well as summary indices of civic and social participation. Because of the difficulty involved in collecting large-scale measures of social capital, few existing datasets lend themselves to a systematic analysis at the local level. A particularly notable exception comes from the Social Capital Project of the Senate’s Joint Economic Committee \cite{JEC2018}, which estimated county-level social capital measures across multiple dimensions. 

We examine the extent to which the geographic patterns revealed by this dataset are also found in online interactions facilitated by local Facebook groups. Our analysis focuses on ``on-platform'' indicators: 42 aggregated and de-identified county-level metrics of Facebook Group usage. These include, among others, the share of users in the county who are in small/medium/large and private/public groups, proportion of private groups observed in the county, whether admins require membership or post approval, etc. We perform exploratory factor analysis on these measures. We then examine the correlations between the computed factors and offline social capital estimates. Finally, we investigate the relationship between the online group factors and measures of community-level public health and inequality outcomes which are mediated by social capital. 

\paragraph{Ethics statement} While the research was carried out, the authors were either employees or contractors of Meta Inc., the owner of the Facebook social networking platform, which also provided the funding and resources for this work. The principal benefit of the analysis is a better understanding of social capital. As detailed in the literature review, social capital has been recognized as a crucial factor for the health of local communities. The dataset described here provides a uniquely rich insight into the deeper structures of social capital, especially as they unfold on increasingly-important online platforms. The most important risk identified relates to breaches of user privacy. To minimize this risk we (1) look at user location information coarsened to the county level and (2) aggregate all computed indicators across all groups local to a county. No individual-level data was used in the study and the dataset is limited to counties with at least 100 users contributing to the aggregates. The authors confirm having read the AAAI Code of Ethics and Conduct and their commitment to abide by it.

\section{Related Work}
Prior work has examined the positive association between individuals' activity on online social networking sites (SNS) and social capital~\cite{ellison2007benefits,vanden2018does,steinfield_social_2008,vanden2018does,tiwari2019social}, though increases in social capital depend on the type of use~\cite{burke2011social}. These studies focused on perceived social capital at the individual level, by asking participants whether they had online contacts they could turn to for various needs.

At the community level, however, studies investigating the connection between online interactions and social capital have typically been limited to individual localized communities, such as a college~\cite{ellison2006spatially} or the ``Netville'' neighborhood that was given broadband internet access earlier than others~\cite{wellman2002networked}. This line of work revealed important conclusions applicable at the individual level. For instance, local online interactions were found to be associated with knowing more individuals in one's community and having higher bridging social capital. Our work aims to shift the focus from individual-level relationships to entire communities. Doing so is only possible when correlating local social capital measures with online activity across a large geographic scale, a possibility afforded to us by the examination of Facebook groups across the United States.

Social capital has come to refer to a particularly immediate understanding of the institutional milieu, a broad set of factors including but not limited to: interpersonal connections, reciprocity, trust and trustworthiness, and shared norms and identity~\cite{lochner1999social}. Higher social capital allows communities to mobilize effectively in the face of crises and makes it possible to solve collective action problems. Such communities fare better in a multitude of societal outcomes such as health, safety, and resilience. In Putnam’s characterization, social capital greases the wheels that allow communities to advance smoothly \cite[p. 288]{putnam2000bowling}. 

The exact meso-level mechanism through which the loose set of concepts termed to be social capital impacts collective outcomes remains an open area of debate.
Transaction costs economics \cite{williamson2008transaction} holds that societies with high social capital have lower transaction costs in everyday business and social life. A differing view is held by 
sociological neo-institutionalists \cite{granovetter1985economic}, who identify in social capital a deeper set of shared understandings which is not directly reducible to interactional cost accounting. Regardless of its exact mechanism of action, there is ample evidence of differences in social capital across countries, as evidenced for instance in comparative studies of international datasets such as the World Values Survey \cite{minkov2012hofstede,  bjornskov2007determinants}. 

Trust is an important aspect to the functioning of online communities as well as social capital in communities in general. Diversity within a geographic community was shown by Putnam to be associated with lower trust between groups and within groups in a large survey conducted in 2000~\cite{putnam07diversity}. Generally, people place more trust in smaller groups~\cite{la2016small}. In the online setting, Ma et al.~\citeyearpar{ma2019people} found people to place more trust in Facebook groups which were smaller, private as opposed to public, having denser networks of friendships within the group, as well as greater age and gender homogeneity. These findings were also supported by interviews of participants in ``Mom-to-mom'' for-sale groups on Facebook, who built trust in private groups of similar members and active group admin involvement. Trust in groups correlated with individuals' propensity to trust and feelings of receiving social support from others in general. Interestingly, increasing participation corresponded to subsequent increases in feelings of trust for the group, suggesting that online engagement may contribute to building social capital~\cite{iyer2020does}.

Some characteristics of online friendship ties, when aggregated to the county level, have been found to correlate with offline social capital. For example, Bailey et al.~\citeyearpar{bailey2018social} found counties with a higher proportion of Facebook ties to friends living more than 100 miles away have higher measures of social capital, and other variables such as income, high school graduation, life expectancy, and social mobility. However, community-level associations, such as the structure and participation in online groups, may be more directly informative about the formation and activation of {\it community} social capital in a location. This is the subject of our present study.

\section{Data and Methods}

We examine the association between community-level social capital by analyzing a dataset that summarizes the activity of US Facebook groups at county level. This data set is available online\footnote{https://doi.org/10.7910/DVN/OYQVEP} \cite{herdagdelen2023}.

For the purpose of this analysis, we consider a group to be active if it has (1) at least one comment, like, or posting event in the 28 days preceding the cutoff date, and (2) at least 10 active members. We consider a member to be active if they had taken at least one action (commenting, posting, liking, etc.) in the group in the 28 days preceding the analysis date. When we compute county-level aggregate statistics over group memberships, we only consider active members. We consider a group a ``US group" if at least 90\% of its members are in the US. For the aggregated age and gender distribution metrics we use self-reported values by the users. Location data is based on predicted city of the users. These values may contain inaccuracies due to self-reports and prediction errors. This is a limitation of our measurements.

\subsection{Defining group locality} 
We take a pragmatic approach and use the intuition that a ``local group'' is composed of people who live near one another, regardless of its function or utility.

A ``local'' Facebook group may be directly tied to a local organization (e.g., a school group), a physical place or institution (e.g., friends of a library), interest group (neighborhood watch, local for sale groups), or just happen to bring people together living in the same area. We consider the mechanism through which a group is salient to a particular location beyond the scope of our study.

As an aggregate measure of locality of a group $g$, we use the probability of two randomly-chosen members of the group being in the same county $c$, or $\lambda_g = P(C_i = C_j)$ where $C_i$ ($C_j$) is a random variable indicating the county associated with the member $i$ ($j$) of the group $g$, with $i \neq j$. This locality measure also known as Simpson index is computed as:

$$ \lambda_g = P(C_i = C_j) =  \frac{\sum_c{n_c(n_c-1)}}{N(N-1)}, $$
where $n_c$ is the number of members of $g$ in county $c$, while $N$ is the total number of members of group $g$. At one extreme, if all members of a group live in the same county the measure is 1. If each member lives in a different county, the measure would take its minimum value 0. If half of the members live in the same county and all other members live in isolated counties, the Simpson Index of the group would converge to 0.25 in asymptotically as $N$ grows. 

\begin{figure}[hbt]
    \centering
    \vspace{-6pt}
    \includegraphics[width=\columnwidth]{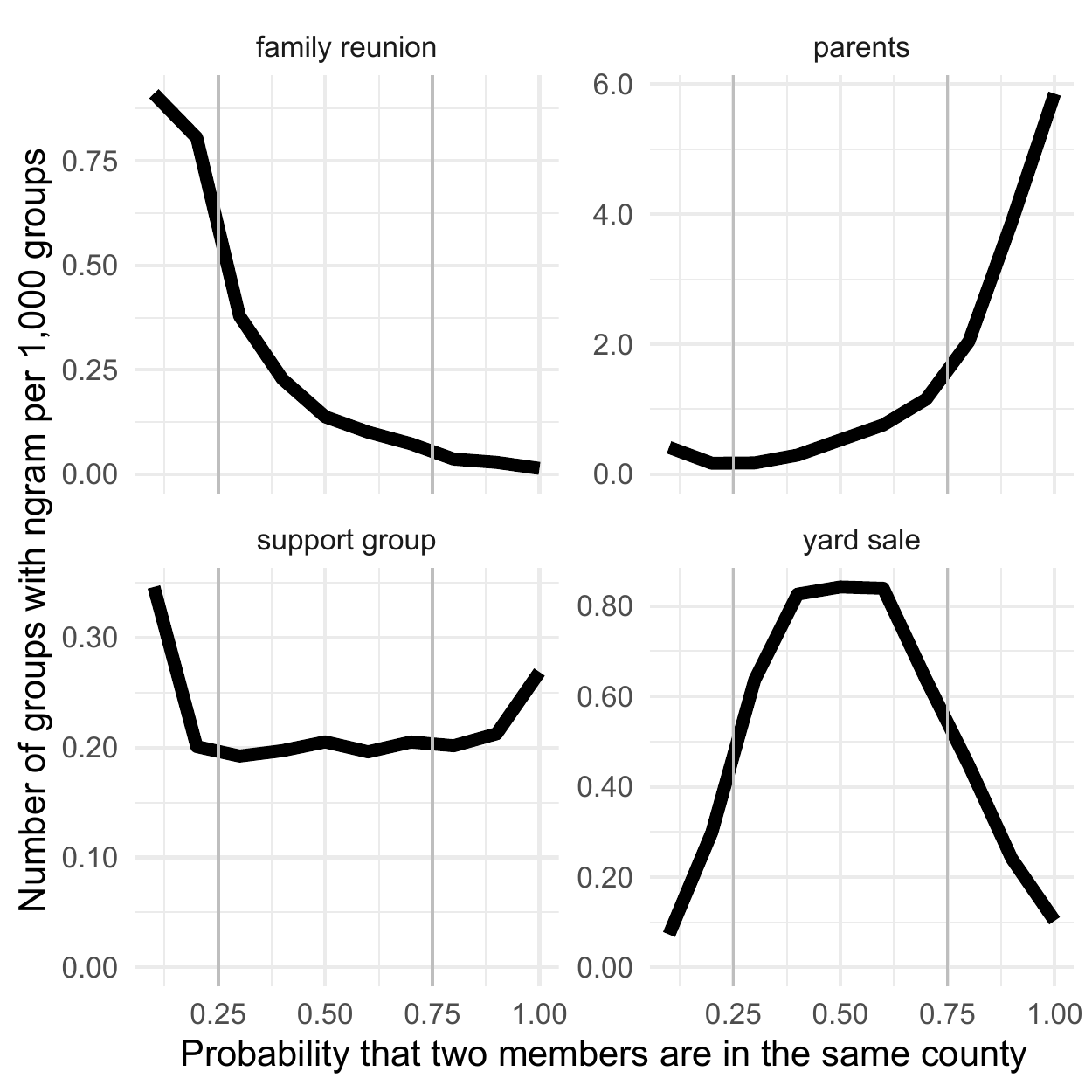}
    \vspace{-12pt}
    \caption{Observation probabilities of illustrative n-grams as a function of group locality score.}
    \vspace{-12pt}
    \label{fig:locality_scores_1}
\end{figure}

As a validation of the locality measure, we looked at probabilities of observing various uni- and bi-grams in public and non-hidden private group names, as a function of locality measure.
In Figure~\ref{fig:locality_scores_1}, we see the relation between the observation probabilities of four illustrative n-grams.

Groups that mention ``family reunion'' tend to be non-local, hinting at the fact that families that reunite tend to live far apart. Groups that mention ``parents'' in the name tend to be very local. ``Yard sale'' groups are more likely have mid-level locality scores suggesting that these kinds of buy and sell groups are of intermediate locality, spanning several counties. The n-gram "support group" illustrates the bimodal nature of support groups; they are more frequently either very local, or very non-local.

Informed by this anecdotal evidence, we decided to set two locality score thresholds: 0.25 and 0.75, dividing groups into three buckets. A group is called \emph{very local} if the probability of two random members being in the same county is above 0.75, \emph{local} if the same probability is between 0.25 and 0.75, and \emph{non-local} if the estimated probability is lower than 0.25. With these thresholds, any group that contains a majority of their members in one county gets classified as local, and if an overwhelming majority lives in the same county it's classified as very local, allowing us to differentiate between groups that are local but still attract memberships from neighboring counties and groups that are strictly localized in one county.

\textbf{Membership in local groups.}  When aggregating local group membership to the county level, we consider the county of the member, even if the group is local to a different county. For example, if a user living in county $A$ is a member of a group that has over 50\% of its members in county $B$, this would still count as a ``local" membership for county $A$.  This broader definition captures participation that is still limited to groups where a majority of members belong to the same county, but allows a group to have a more flexible local ``area of influence", even if it extends across sometimes arbitrary-seeming county borders.

\subsection{Group size}

We consider two group-level characteristics to have particular relevance to the relationship between local Facebook groups and community-level social capital. We expect group dynamics to vary as group size increases and group members are less likely to be directly connected through social relations. Given the extent to which trust is constitutive of social capital, we likewise expect important interactions between group privacy levels and social capital.

We partitioned groups according to their membership size $m$ into four buckets: $m < 30$ is very small, $30 \leq m < 100$ is small, $100 \leq m < 1000$ is mid-size, and groups with at least 1,000 members are categorized as large. The strategy of dividing groups into progressively-larger size buckets was chosen due to the heavy-tailed nature of the group size distribution.  

\subsection{Group Privacy}

Facebook groups have different privacy levels that determine the who can see the group content and membership list. \textit{Public} groups have the highest visibility and both the content and member list are visible to non-members. Content in \textit{Private} groups is only visible to members. Furthermore, private groups can be \textit{hidden}, meaning that the groups itself is only visible to its members and can be joined by invitation. 

\subsection{Offline Social Capital Estimates}
Efforts to measure social capital systematically and at scale have been comparatively rare. This dearth of information is particularly pronounced for additional requirements of recency and geographic details -- both necessities for providing a detailed picture of the current state of social capital in the United States.

 The \citet{JEC2018} of the Senate’s Joint Economic Committee (JEC) offers the most detailed and timely estimate of social capital we were able to identify at the time of the study. In 2018, this project released a comprehensive dataset of multiple social, economic and physical indicators, aggregated at the level of individual US counties. Some of these indicators are used as components of a composite social capital estimate and some are used as benchmarks against which the social capital estimate is evaluated in terms of predictive power. Given our theoretical interest in associational social capital, we focus on the \textit{community health} components of the Social Capital Project. 

\textbf{Community Health} is a composite index based on the number of county-level non-religious non-profits per person, religious congregations per person, and state-level data on percentage of people who attended public meetings, report working with neighbors to solve common problems, participate in volunteer work, etc. This is one subdimension used in the JEC study, but since it captures the associational activity in a county, we decided to use it as our dependent variable.

\begin{figure}
    \centering
    \includegraphics[width=\columnwidth]{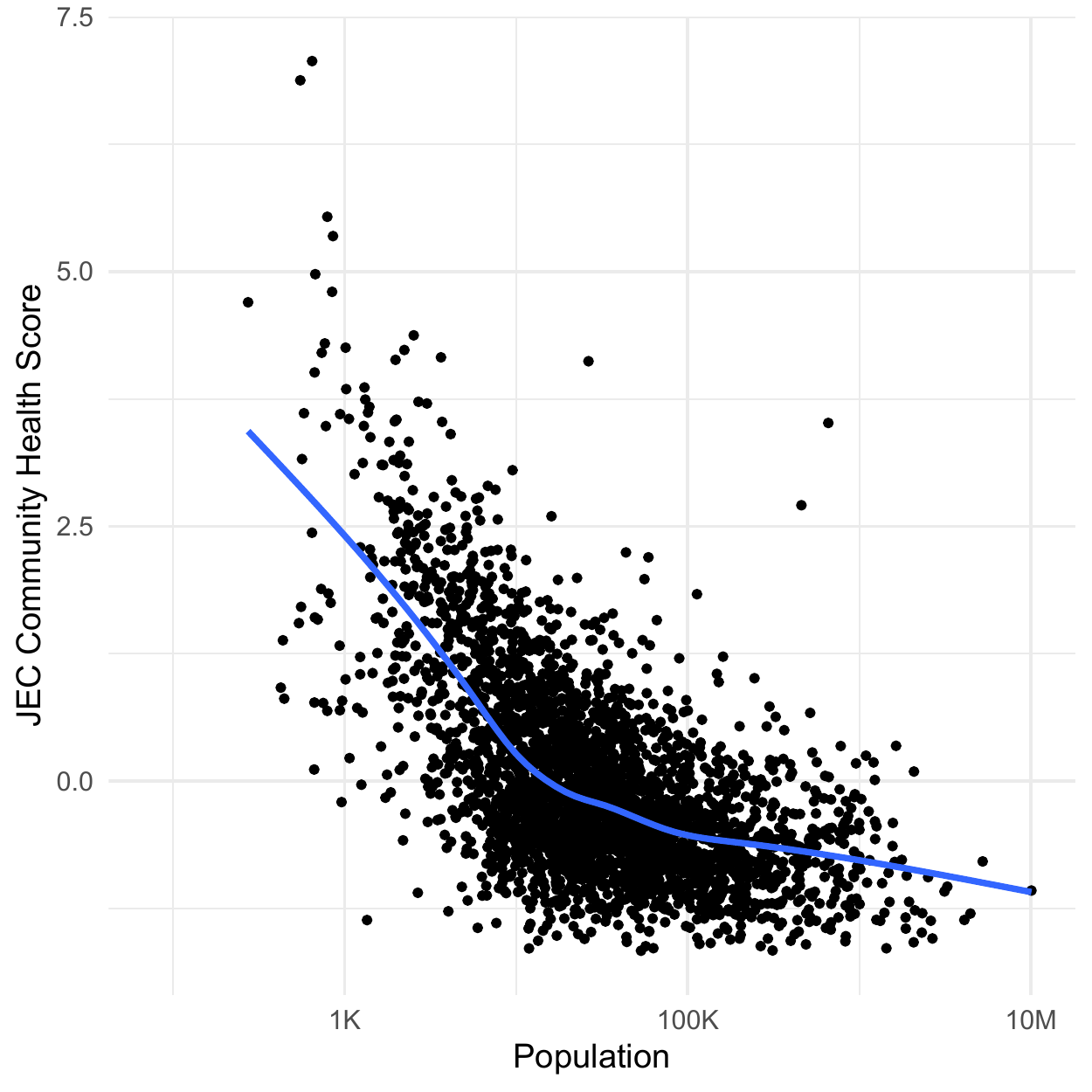}
    \vspace{-12pt}
    \caption{Community Health Indices of US counties, as a function of county population. Spearman correlation between the two is -0.58.}
    \label{fig:population_vs_community}
    \vspace{-18pt}
\end{figure}

As we see in the scatter plot of Figure~\ref{fig:population_vs_community}, counties with higher population tend to have lower values of the community health index, with an overall correlation of (-0.58). While the implications of this relation are beyond the scope of this study, interested readers can refer to the literature on scaling effects of populations and organizations and resources~\cite{gastner2006optimal, bettencourt2007growth}. 

We also compute a community health index adjusted for population and density by fitting an OLS with the logarithm of these values (and their squares) as the only covariates and using the residuals as the adjusted community health index. In other words,

$$y_i=\beta_1 log(p) + \beta_2 log(p)^2 + \beta_3 log(d) + \beta_4 log(d)^2 + \epsilon_i$$ 

where $p$ and $d$ are total population and population density of county $i$, respectively, with $y_i$ the community health index of county $i$. We use the residual $\epsilon_i$ as the adjusted index. We will use both unadjusted and adjusted indices in our exploratory analyses. 

\textbf{Generalized Trust} (occasionally called ``social trust'') is one of the more common operationalizations of social capital \citep{newton2001trust}. Broadly speaking, it is defined as the willingness to trust a generic alter, or ``the potential readiness of citizens to cooperate with each other and [the] abstract preparedness to engage in civic endeavors with each other'' \citep{stolle2002trusting}. Generalized trust is typically measured attitudinally, at the individual level -- the relevance of this indicator being bolstered by the inclusion of a generalized trust question in the World Values Survey. The work of \citet{bjornskov2007determinants} reveals striking country-level differences for this construct. In the United States, strong inter-regional differences were documented by \citet{simpson2006poverty}. Recently, \citet{wu2020does} even established the persistence of discrepancies in generalized trust attitudes for individuals moving between regions with varying levels of generalized trust.

Generalized trust is arguably uniquely suited as an individual-level construct that can nonetheless reveal deep, community-level differences in social capital. Percentage of people living in an area who exhibit high levels of trust is used as a community-level indicator. Unfortunately, a county-level generalized trust dataset against which one could compare this indicator is lacking.\footnote{This gap in existing measurement was also identified by the authors of the afore-mentioned JEC study \citep[][p.19]{JEC2018}}. State-level generalized trust measurements are however available thanks to the Generalized Social Survey \cite{Neville_2012}. We rely on these measures as the best available indicators of social trust for local geographies in the United States. 

\subsection{Group-Based Indicators}

We compute 42 aggregated and de-identified county-level indicators of Facebook Groups usage. These indicators can be grouped under two categories:

\subsubsection{Participation in different group types} 
In this category we measure what percentage of active Facebook users in a county participate in different group types by size, privacy, and locality. First, we partition each group into one of 36 mutually exclusive buckets, organized along three dimensions based on privacy (public / private / hidden), number of users in the group (very small, small, mid-sized, large), and locality (very local, local, non-local). Then we compute the percentage of users living in a county who participate actively as a member in at least one such group. For instance, the indicator (private, very local, very small) for a county represents the percentage of Facebook users who live in the county and who are active members of at least one very small, private group with a very high locality score. 

\subsubsection{Aggregate local group characteristics} 
Here we aggregate the characteristics of local groups observed in a county. We only use groups where the majority of members live in the same county (i.e., locality score >= 0.25) and include a group in the averaging only for the county where the majority of members live in\footnote{We also used a less restricted way of incorporating a group in averaging for a county, using the ratio of group members in the county as a weight. The result did not change meaningfully.}. The characteristics are:

\begin{itemize}

\item \emph{Content gating}. Ratio of local groups in a county where admin approval is required for posts before they become visible to other members.

\item \emph{Member gating}. Ratio of local groups in a county where the group has at least one of three controls for joining as a new member: admin approval, agreeing to group rules, or questions that are required to be answered for joining.

\item \emph{Mean gender diversity}. We use the probability that two randomly chosen members of a group having different genders as the gender diversity of the group. This measure is the complement of Simpson's index for gender distribution (also known as the Blau score). Formally, gender diversity of a given group $g$ is $\gamma_g=1 - P(G_i=G_j)$ where $G_i$ ($G_j$) is a random variable indicating the self-reported gender (male or female) of a member $i$ ($j$) of the group $g$, with $i \neq j$.

The average Blau score of local groups observed in a county $c$ is used as the county-level gender diversity indicator for the county:

$$\frac{\sum_{g\in G_c}\gamma_g}{|G_c|},$$
where $G_c$ is the set of all local groups in county $c$, and $\gamma_g$ is the group-level gender diversity as defined above.

\item \emph{Mean locale diversity}. We use the probability that two randomly chosen members of a group having different locale setting for Facebook as the locale diversity of the group. Example locales include ``English (US)'' or ``Portugês (Brasil)." Average locale diversity scores that are local to a county is used as the mean locale diversity indicator for the county. Formally, this is defined in a similar manner to the gender diversity defined above,

$$\frac{\sum_{g\in G_c}\mu_g}{|G_c|},$$
where $\mu_g=1 - P(M_i=M_j)$, giving us the probability that locale settings of two randomly chosen members of $g$ are different.

\item \emph{Mean age diversity}. Average inter-quartile range of the age distribution of local groups. The inter-quartile range is defined as the number of years between the 25th and 75th percentile values of the empirical age distribution.

\item \emph{Mean tie density}. Ratio of actually realized (made) Facebook friendship ties between group members to all possible pairs.

\end{itemize}

High colinearity among our indicators renders standard regression techniques inappropriate for exploratory data analysis. We could collapse these indicators, using marginal participation rates along salient dimensions, instead of the joint measurements along privacy, size, and locality. However, in the absence of a strong theoretical basis and \textit{a priori} expectations of which indicators go together, we decided to employ exploratory factor analysis to study the internal correlation structure of online platform indicators and their relation to offline societal benchmark indicators.

\section{Results}

Before we discuss the results in more detail, we illustrate three on-platform indicators to build a stronger intuition of the data we are studying.

\textbf{1. Local group membership prevalence} is defined as the ratio of users who live in a county and who are active members of at least one local or very local group, to all active members in the same county. In the following exploratory analyses, we use sub-components of this indicator, broken down by privacy, size, and locality of the groups, but for illustration purposes here we use the composite indicator.

\begin{figure}[h]
\centering
\includegraphics[width=\linewidth]{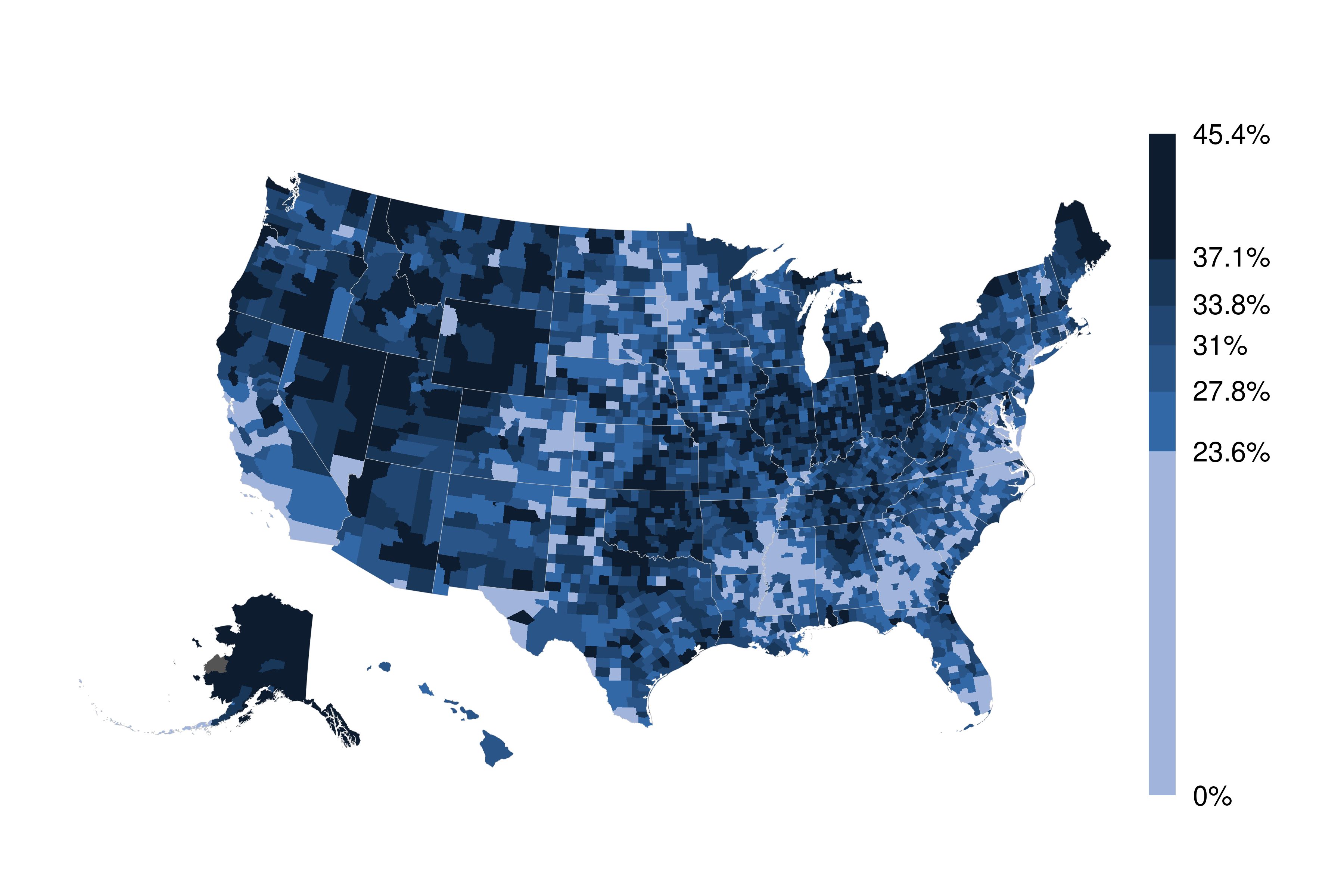}
\vspace{-12pt}
\caption{Share of active local group members (\%) who are in at least one local group all active Facebook users. Each distinct color represents an equal-sized (in terms of counties) bucket.}
\label{fig:local_group_participation}
\vspace{-6pt}
\end{figure}

A county-level choropleth (Figure~\ref{fig:local_group_participation}) reveals strong spatial autocorrelation and large-scale patterns in local group membership prevalence. In certain areas of the Midwest, Great Plains, Coastal California, and particularly the Deep South, we observe low levels of local group membership prevalence.

Intuitively, local group membership prevalence may be expected to reflect associational activity and social capital in an area. However, we do not observe a correlation ($\rho= 0.04$) between \emph{\% local users} and \emph{community health}, the JEC index measuring associational activity in a county, such as volunteering, attending public meetings, working with neighbors to fix things, etc. 

\begin{figure}[h]
\centering
 \includegraphics[width=\linewidth]{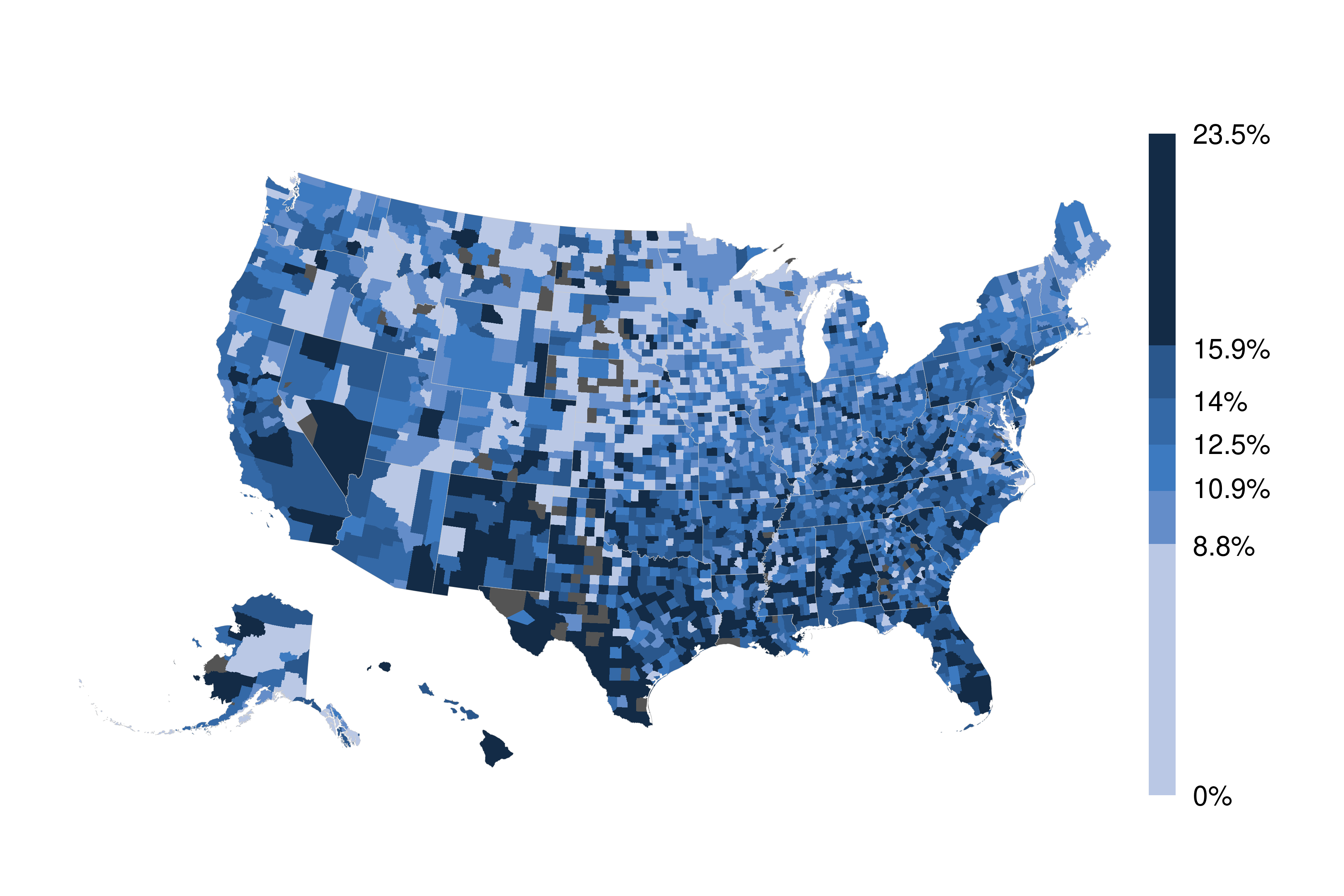}
 \vspace{-12pt}
\caption{Percentage of groups in each county that require admin approval for contents posted to the group.}
\label{fig:content_gating}
\end{figure}

\textbf{2. Content control}
The second example indicator is \emph{p(content gated)}, the percentage of local groups that do not allow members to post content without admin approval, shown in  Figure~\ref{fig:content_gating}. The choropleth reveals a gradient of increasing content control from North to South. The lowest levels of the indicator are observed in parts of the Midwestern United States (North Dakota, South Dakota, Minnesota, Iowa, and Wisconsin). The highest levels of the indicator are observed in the Southern States (Louisiana, Mississippi, Alabama, Georgia, South Carolina, Texas, and Florida).

\begin{figure}[h]
\centering
\includegraphics[width=\linewidth]{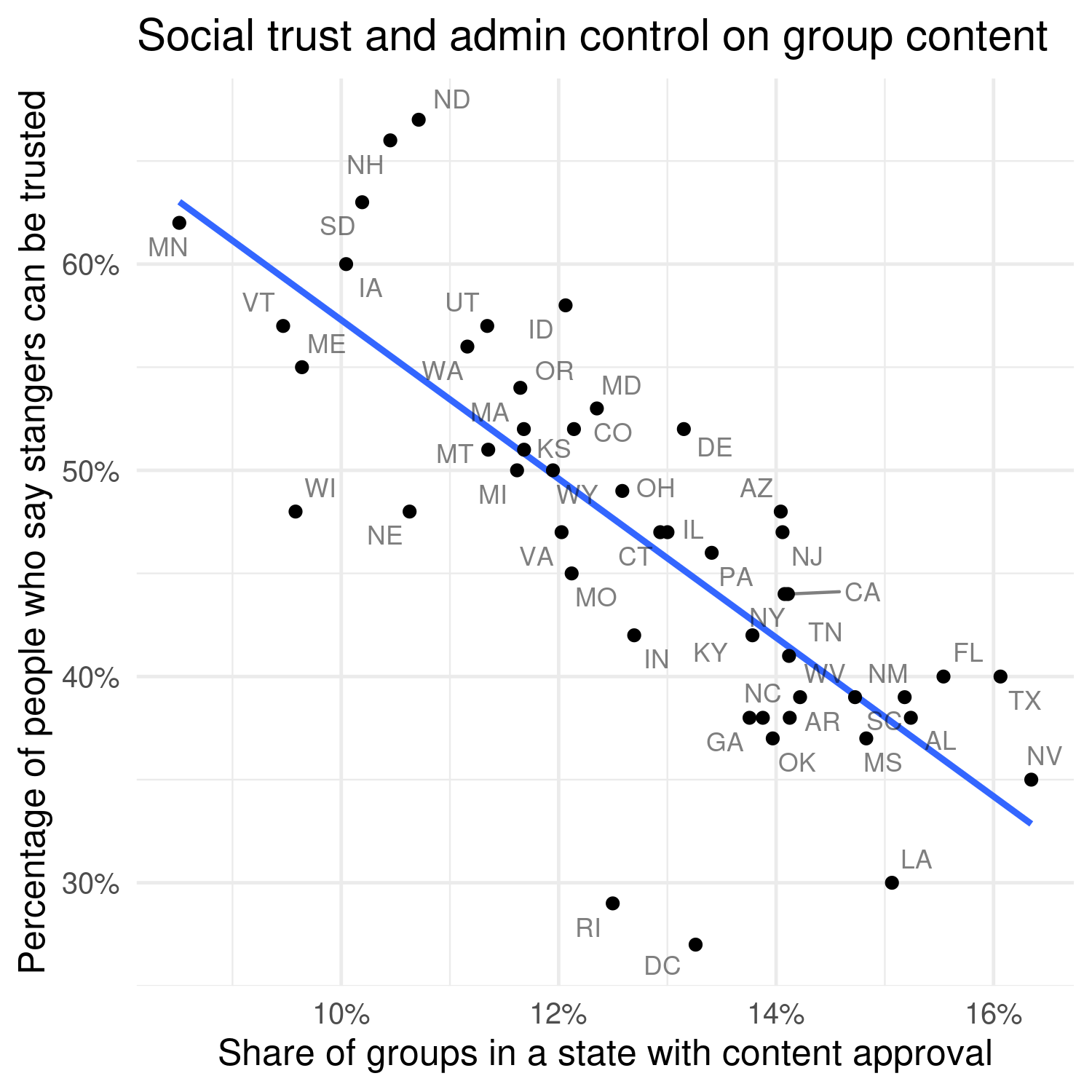}
\vspace{-12pt}
\caption{Generalized trust (in strangers) and share of groups \emph{without} admin approval are strongly correlated ($r=0.81$).}
\label{fig:state_trust_gating_scatter}
\vspace{-12pt}
\end{figure}

We aggregated the data at the state level to represent what \% of local groups require admin approval. Doing so allowed us to compare our findings against the previously-discussed trust questions in the General Social Survey~\citep{Neville_2012}. We observe a strong negative relation ($\rho=- 0.81$) between the share of groups with admin approval and the \% of survey respondents who agreed that strangers can be trusted.  \emph{p(content gated)} also has similarly strong correlation ($\rho = -0.82$) to the trust in neighbors measure provided by the ~\citet{JEC2018}. A scatter plot of both measures depicting the strong bivariate relation is given in Figure~\ref{fig:state_trust_gating_scatter}.

However, such correspondence is not shown by all group-derived indicators that one might hypothesize reflect trust. For example \emph{p(member gated)}, the probability that a group places controls on who can become a member of the group, is moderately correlated with trust in neighbors ($-0.56$), but exhibits only a weak correlation with generalized trust ($-0.25$). 

While one could come up with plausible \textit{post-hoc} explanations for why the above results make sense, our conclusion is not to trust our hypotheses about single indicators, despite their plausibility. Instead, we use a data-driven approach to unearth latent factors that all of our indicators measure collectively. In the next section, we provide the results of exploratory factor analysis. 

\begin{figure}[htb]
\centering
\includegraphics[width=\linewidth]{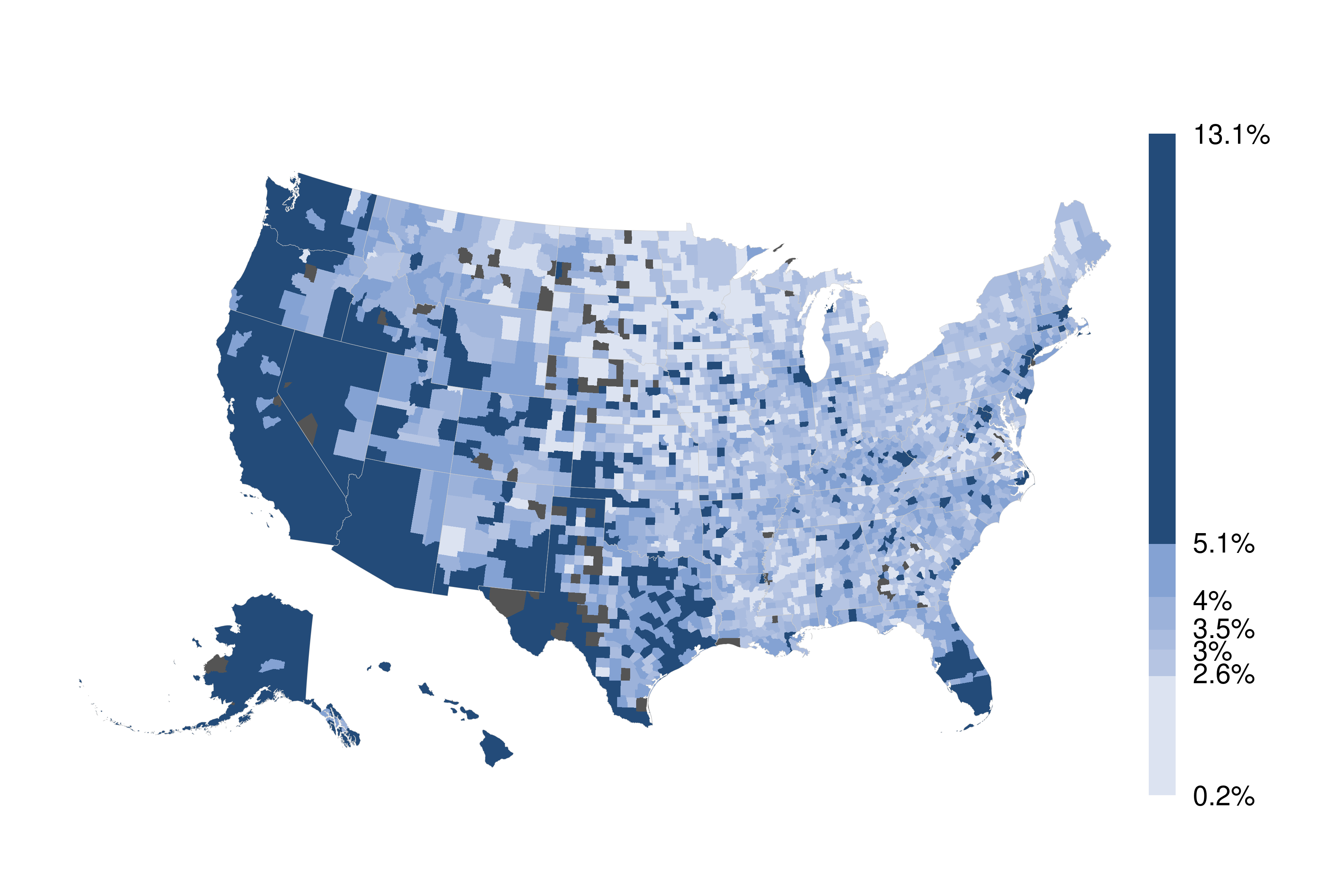}
\vspace{-12pt}
\caption{Average locale diversity of local groups observed in counties.}
\label{fig:locale_diversity}
\vspace{-12pt}
\end{figure}

\begin{figure}[htb]
\centering
\includegraphics[width=\linewidth]{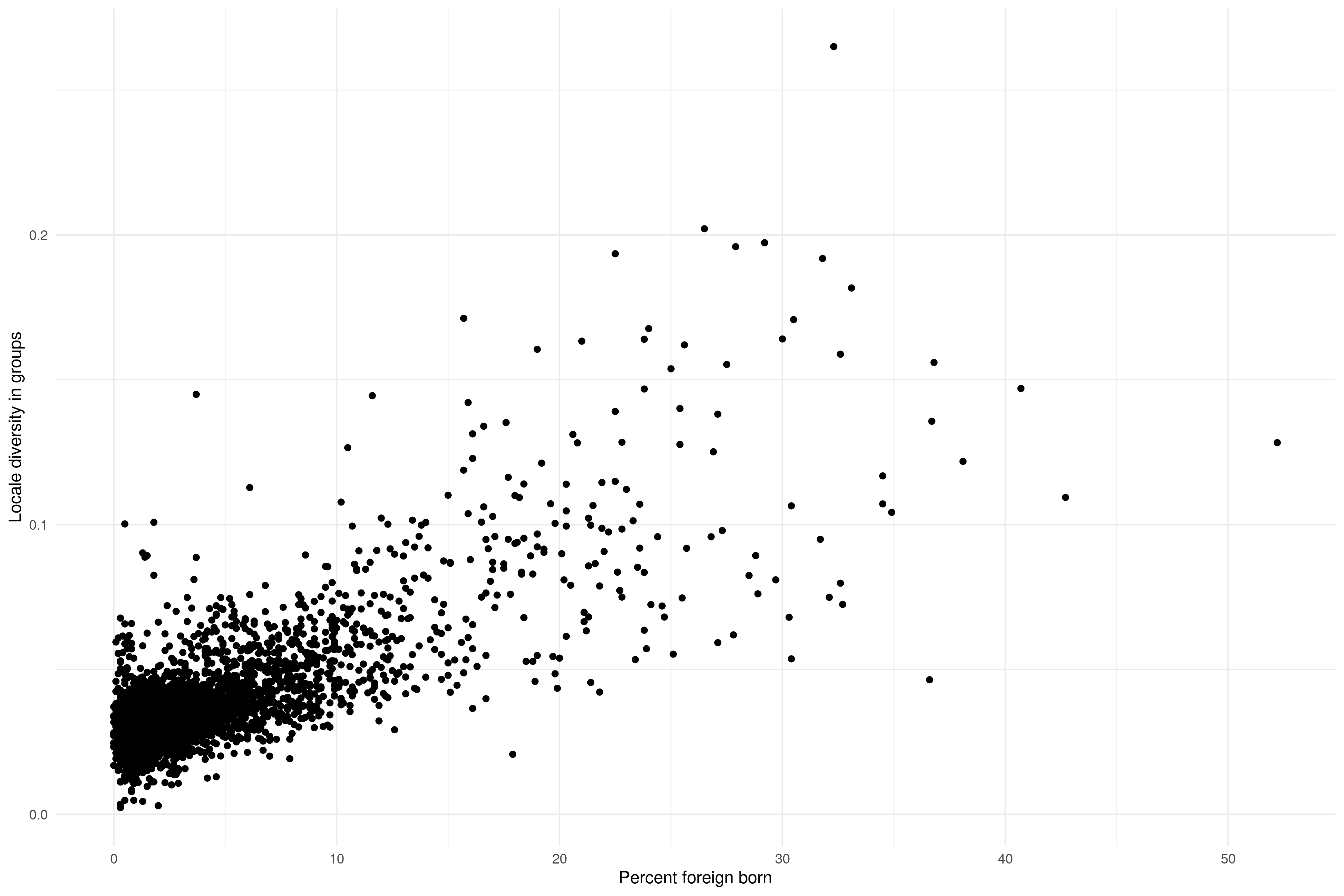}
\caption{Locale diversity and percentage foreign-born population in each county ($\rho=0.61$.)}
\label{fig:locale_diversity_foreign_born_scatter}
\vspace{-18pt}
\end{figure}

\textbf{3. Linguistic diversity}
Figure~\ref{fig:locale_diversity} of the above-defined {\emph mean locale diversity} reveals an unsurprising pattern: Areas that receive in-migration, such as border counties, the West Coast, and other urban areas, have higher levels of locale diversity, reflecting the demographic characteristics of these areas. In Figure~\ref{fig:locale_diversity_foreign_born_scatter}, we plot the locale diversity of the groups in each county with respect to the percentage of foreign-born population and note a strong correlation ($\rho=0.61$).

\subsection{Extracting Latent Factors}

To identify the latent factors that our indicators can collectively explain, we applied exploratory factor analysis (EFA). We used the maximum likelihood method and extracted 4 factors as suggested by a visual inspection of the scree plot (see Appendix for details), using VARIMAX rotation. The factor structure was robust to using OBLIMIN rotation and we obtained qualitatively similar results using a varimax-rotate principal component analysis. Factor loadings are summarized in Figure\ref{fig:factor_loadings}, while loading values are visualized for each county in the Appendix, which also provides explicit factor loadings for all the component variables.

We flipped the signs of the factors in a way that made interpretations easier. The sign selection is essentially an arbitrary decision that does not change numerical results or their interpretation.

\begin{figure}[h]
\centering
\includegraphics[width=\linewidth]{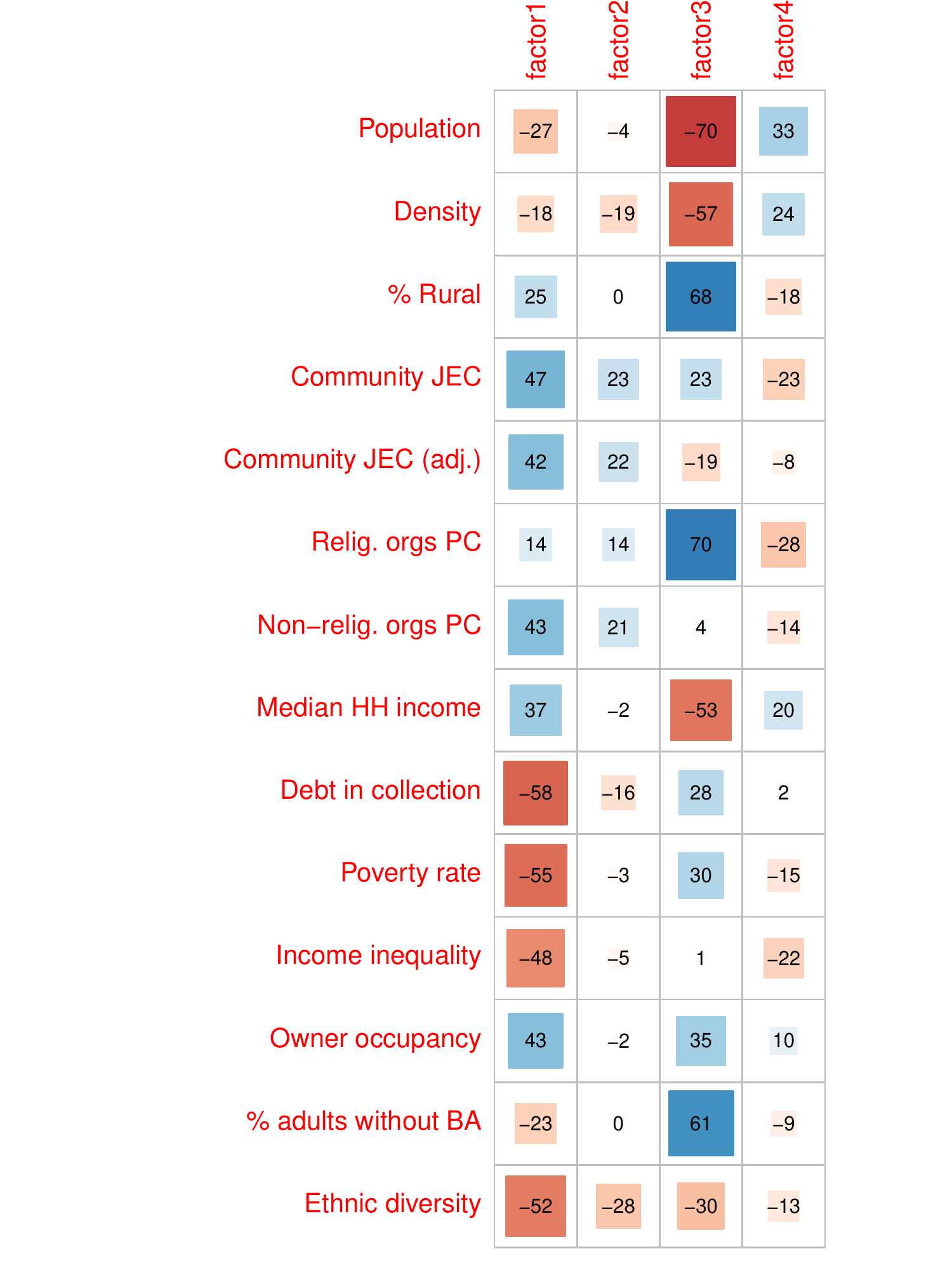}
\vspace{-12pt}
\caption{Spearman correlation between the factors and social indicators.}
\label{fig:factor_cormat}
\vspace{-12pt}
\end{figure}

In the following sections, we follow the same recipe to describe each factor in order. First, we interpret the factor loadings on individual indicators and note common patterns.
Then, we report correlations with population and community health measures. We also take note of relevant socioeconomic and demographic indicators which likewise exhibit relatively strong correlations.
We also present and discuss the geographical patterns over the choropleths of the factors and note pockets of high- and low-value areas.

Finally, we provide a list of group name ngrams that have high salience for each factor: Using the regression factor loadings from the county-level dataset, we compute a ``faux-factor'' score for each group (e.g., if a group is a public, local, small group we use the factor loading of public\_local\_small). Then, we compute the average group factor scores of unigrams and bigrams occurring in group names. The top-scoring ngrams for each factor are identified as salient ngrams.

\begin{figure}[t]
\parbox{\columnwidth}{
\centering
\begin{subfigure}{0.8\columnwidth}
\includegraphics[width=0.8\columnwidth]{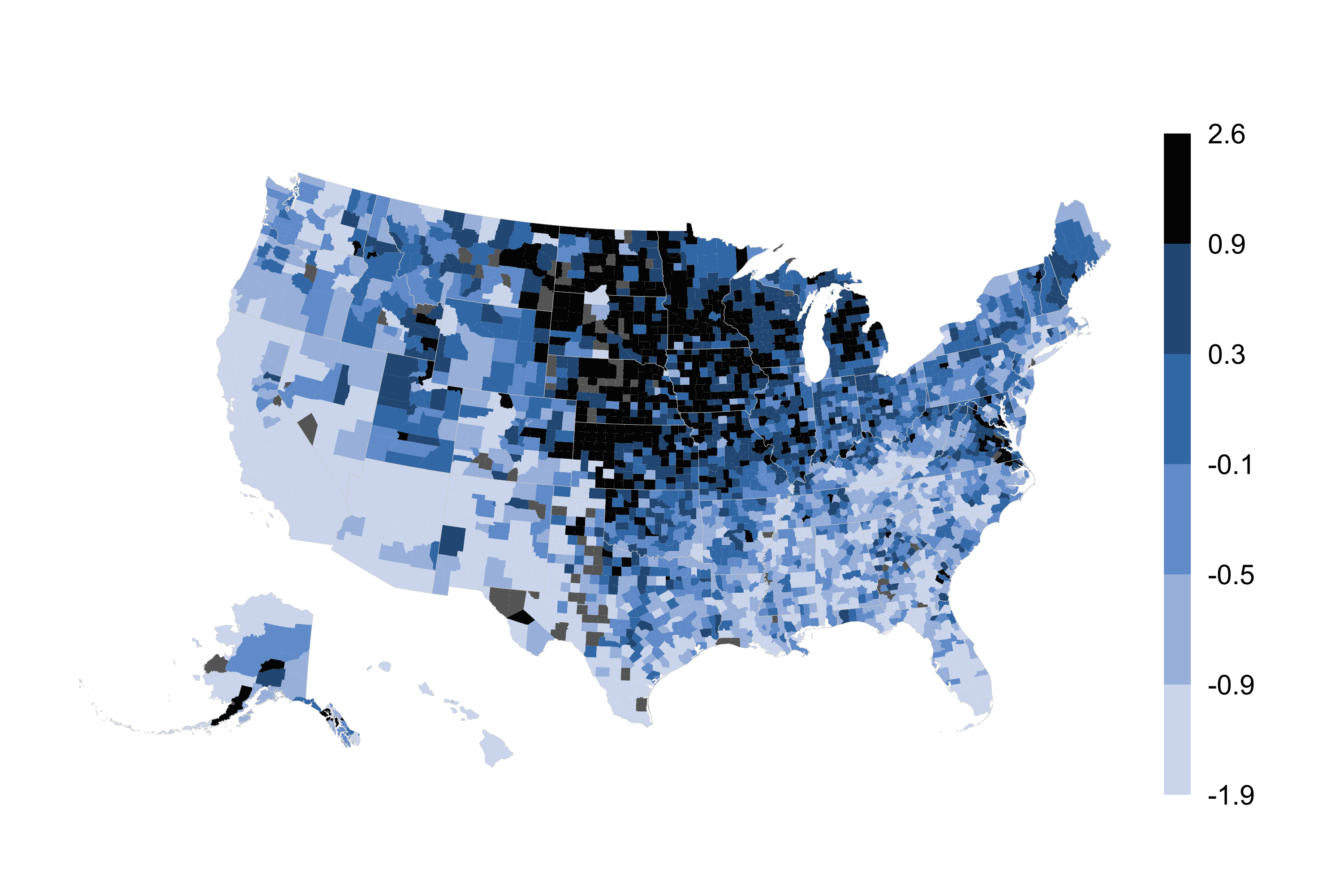}
\vspace{-12pt}
\subcaption{F1: small, tightly-knit, local groups.}
\label{fig:F1_choropleth}
\end{subfigure}
\begin{subfigure}{0.8\columnwidth}
\includegraphics[width=0.8\columnwidth]{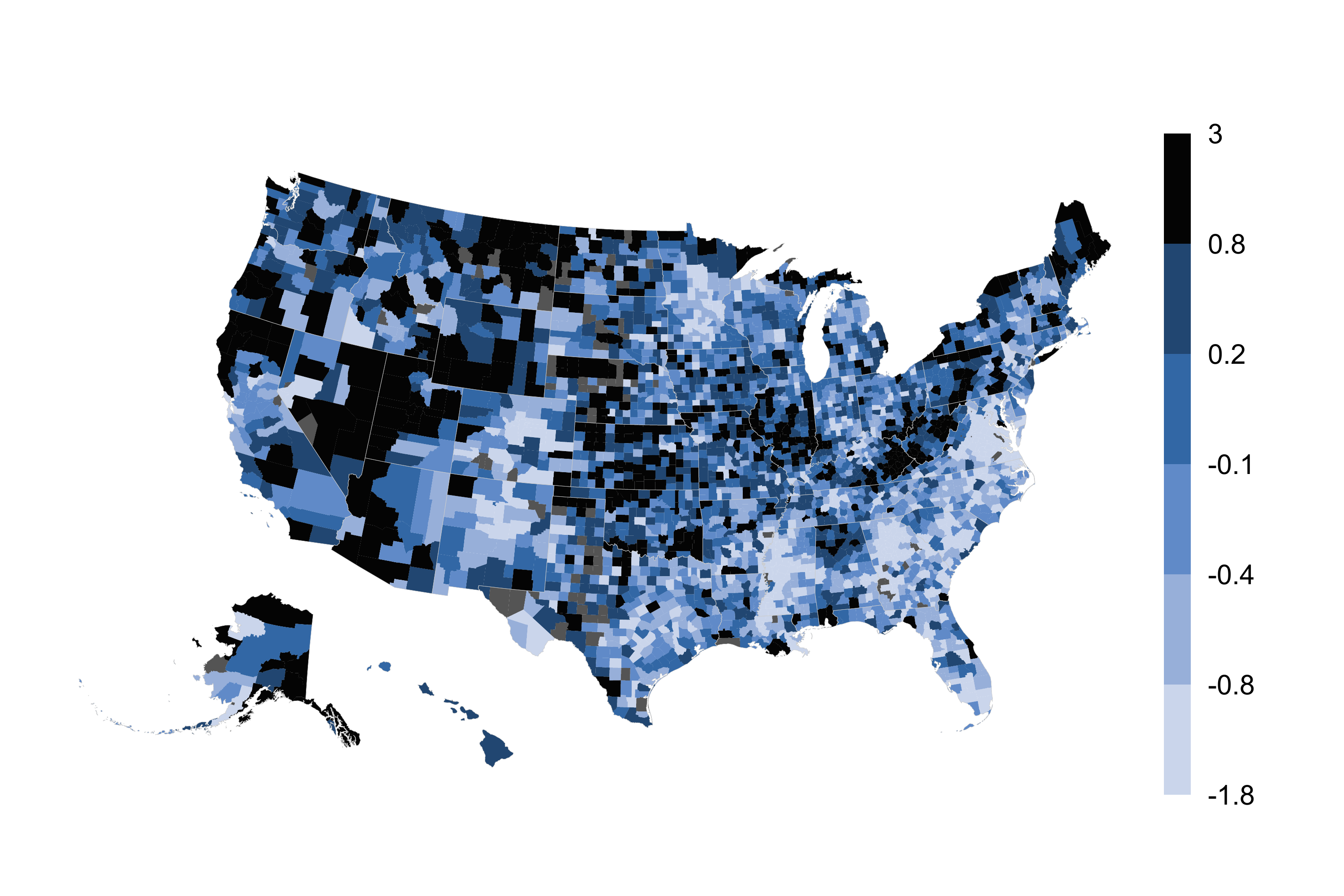}
\vspace{-12pt}
\subcaption{F2: Very local and small groups.}
\label{fig:F2_choropleth}
\end{subfigure}
\begin{subfigure}{0.8\columnwidth}
\includegraphics[width=0.8\columnwidth]{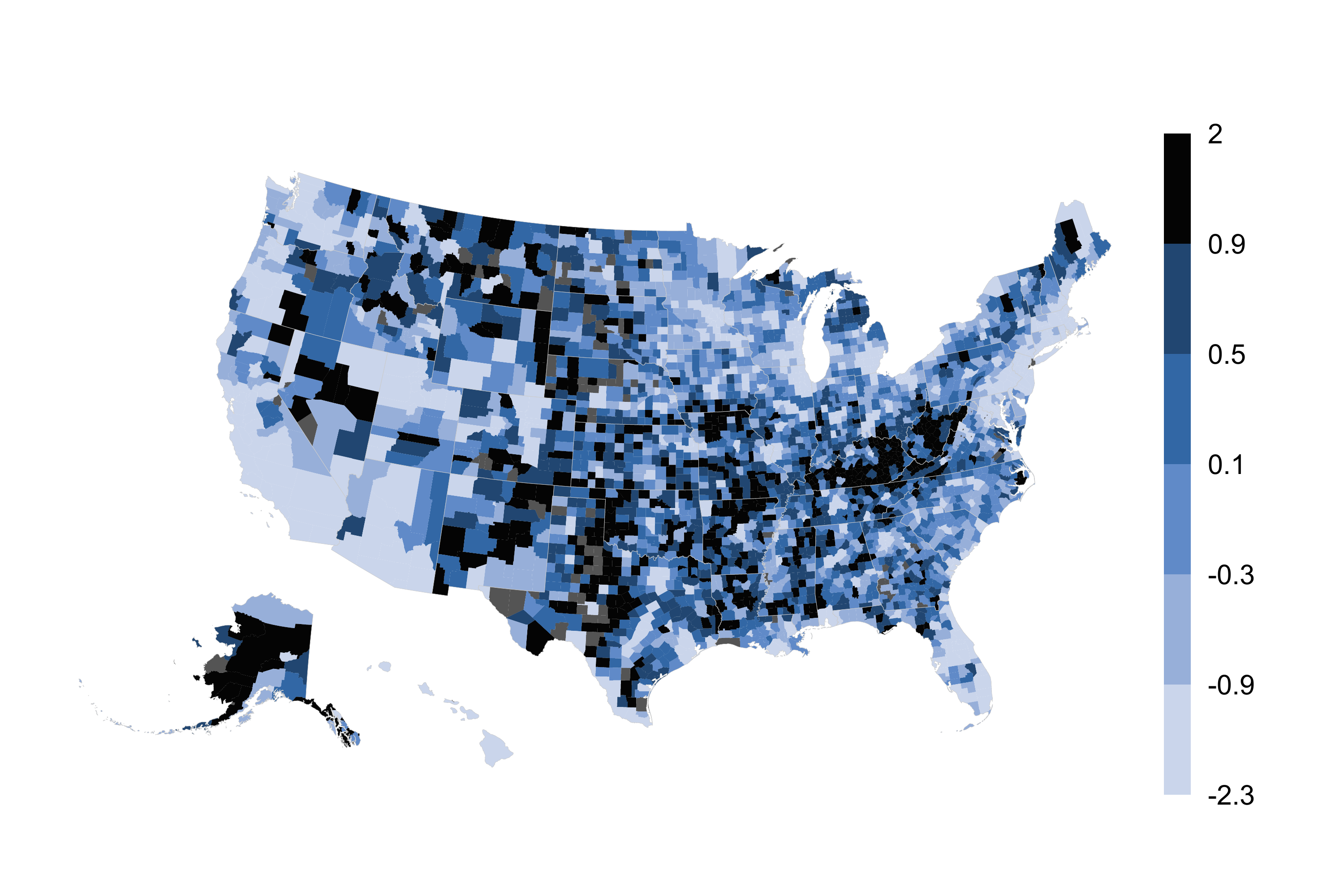}
\vspace{-12pt}
\subcaption{F3: Public, large, age-diverse groups.}
\label{fig:F3_choropleth}
\end{subfigure}
\begin{subfigure}{0.8\columnwidth}
\includegraphics[width=0.8\columnwidth]{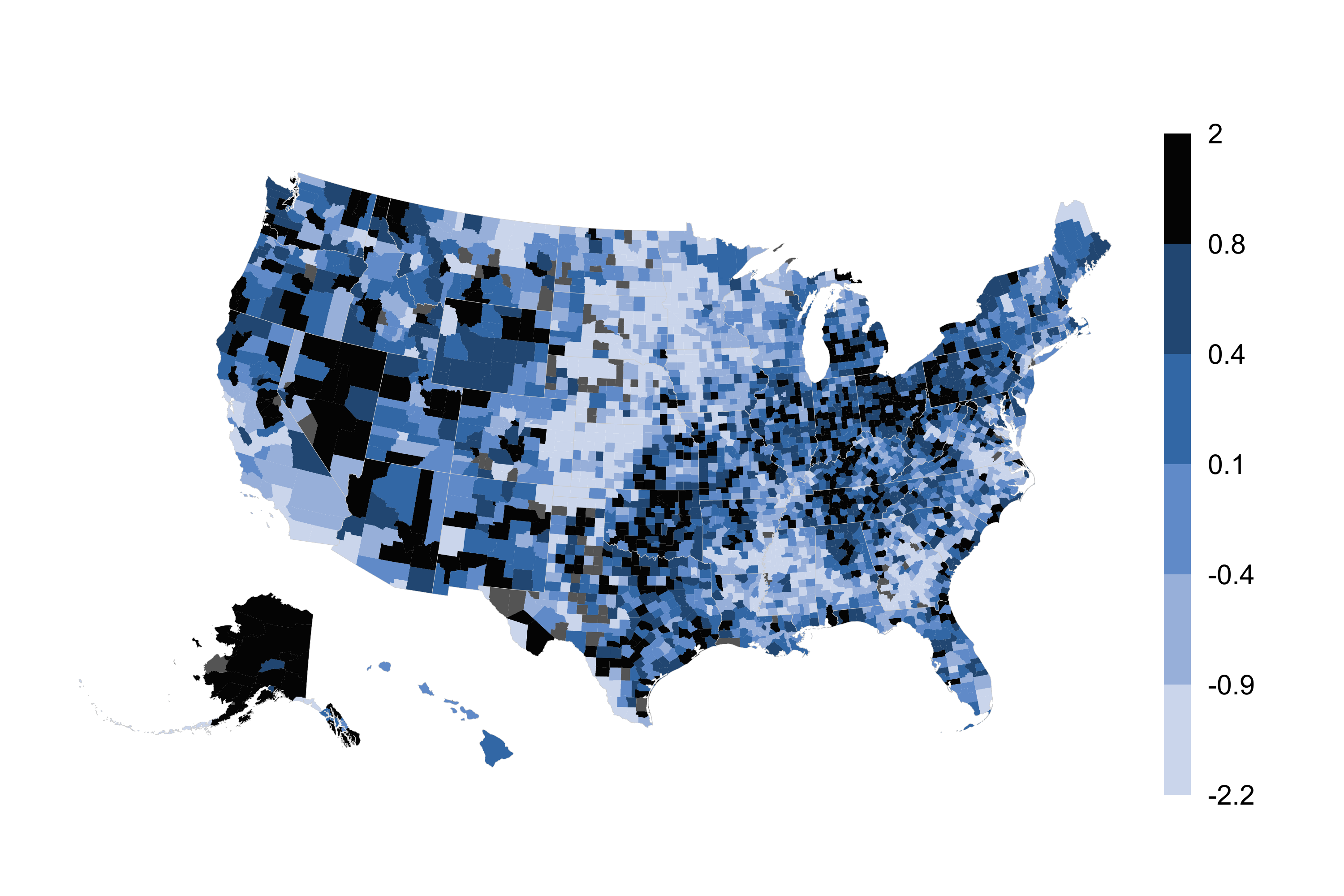}
\vspace{-12pt}
\subcaption{F4: Partially local, member-controlled and bridging groups.}
\label{fig:F4_choropleth}
\end{subfigure}
\caption{Choropleths of factor values at the county level.}
\label{fig:factor_choropleths}
\vspace{-18pt}
}
\end{figure}

\subsubsection{F1: Small, tightly knit, non-local groups}

F1 captures counties with small non-local groups with dense ties, low content and member gating, and low locale diversity, In Fig.~\ref{fig:factor_loadings}, we plot the loading factors of group-participation indicators. This factor's values are highest in the Midwest and Utah, and lowest in California and the Southwest (Figure \ref{fig:factor_choropleths}). 

\begin{figure}[htbp]
\centering
\includegraphics[width=\linewidth]{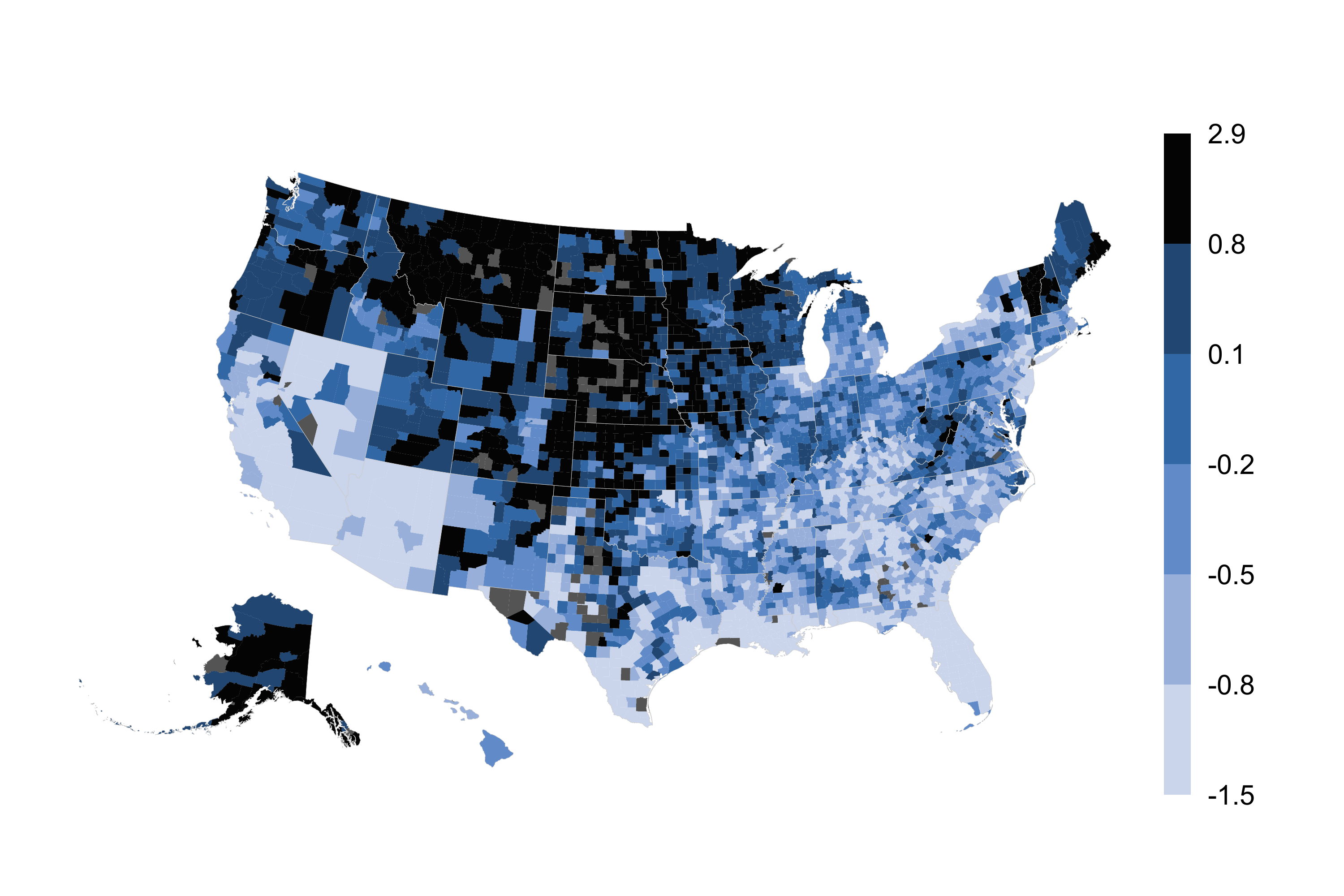}
\vspace{-18pt}
\caption{Choropleth of JEC community health subindex.}
\vspace{-18pt}
\label{fig:jec_choropleth}
\end{figure}

Figure~\ref{fig:factor_cormat} reveals that F1 is correlated to both unadjusted ($\rho=0.47$) and population-adjusted community health ($\rho=0.42$). The geographical distribution is visibly similar to that of community health that we plot in Fig.~\ref{fig:jec_choropleth}. Among the two major components of JEC's measure, F1 is correlated with non-religious, non-profit organizations per capita ($\rho=0.43$), but not with religious organizations per capita ($\rho=0.14$). 

F1 is also positively correlated with economic health indicators, namely, median household income ($\rho=0.37$) and owner-occupied housing ($\rho=0.43$). In addition, F1 is negatively correlated with debt in collection ($\rho=-0.58$, poverty rate ($\rho=-0.55$) and income inequality ($\rho=-0.48$). Finally, it is negatively correlated with ethnic diversity in the county ($\rho = -0.52$).

A detailed analysis of the relationship between diversity and social capital is beyond the scope of our study. However, given the negative associations between F1 and admin control, locale diversity of groups, and ethnic diversity of the county populations, we wanted to provide an additional layer of analysis, inspired by the literature on trust and diversity \cite{simpson2006poverty}. In Table~\ref{tbl:lm_f1_regional}, we provide the coefficient estimates of a linear model with F1 as the dependent variable, focusing on the coefficient estimates of the binary indicator of whether the county is in the Southern United States and ethnic diversity, while controlling for population, percent rural, median household income, high school graduation rate, and religious congregations per capita. The reported coefficient estimates are for standardized input variables (mean subtracted and divided by two standard deviations, following \citet{gelman2008scaling}).

The main effect sizes of \textit{is\_south} (-0.071) and \textit{ethnic diversity} (-0.539) are in line with the correlations we reported above and also previous discussions of diversity and trust and social capital \cite{putnam07diversity}. Higher ethnic diversity and being in the Southern US are associated with lower F1 values. However, the high value of the coefficient of the interaction term \textit{is\_south:ethnic\_diversity} (0.818) suggests that in the Southern United States, the relation between diversity and F1 goes against the direction predicted by trust literature. In the Southern US, counties with more ethnic diversity exhibit higher levels of F1. The interaction term effectively cancels out the negative contribution of increasing ethnic diversity in the South.

\begin{table}[h] \centering 
  \small
\begin{tabularx}{\columnwidth}{l @{\extracolsep{-5pt}} d{2} @{\extracolsep{-5pt}}  d{2}} 
 & \multicolumn{1}{l}{\hspace{5em} \textit{Coef.}} 
 & \multicolumn{1}{l}{\hspace{2em} \textit{S.E.}} \\
\hline \\
 log(population) & -0.440^{***} & (0.041) \\ 
 \% rural & 0.072^{**} & (0.036) \\ 
 median household income & 0.387^{***} & (0.029) \\ 
 \% adults graduated high school & 0.776^{***} & (0.032) \\ 
  religious congregations per 1000 & 0.273^{***} & (0.035) \\ 
  is south & -0.071^{**} & (0.028) \\ 
  ethnic diversity & -0.539^{***} & (0.028) \\ 
  is\ south : ethnic diversity & 0.818^{***} & (0.049) \\ 
 Constant & -0.076^{***} & (0.012) \\ 
\hline \\[-1.8ex] 
Observations & \multicolumn{2}{c}{3,027} \\ 
R$^{2}$ & \multicolumn{2}{c}{0.617} \\ 
Adjusted R$^{2}$ & \multicolumn{2}{c}{0.616} \\ 
Residual Std. Error & \multicolumn{2}{c}{0.599 (df = 3018)} \\ 
F Statistic & \multicolumn{2}{c}{$606.672^{***}$ (df = 8; 3018)} \\ 
\hline 
\hline \\[-1.8ex] 
\multicolumn{1}{l}{$^{*}$p$<$0.1; $^{**}$p$<$0.05; $^{***}$p$<$0.01} \\ 
\end{tabularx} 
\vspace{-6pt}
\caption{Coefficient estimates of a linear model estimating F1. Numerical variables are standardized by centering around 0 and divided by 2 standard deviations.}
\label{tbl:lm_f1_regional}
\vspace{-6pt}
\end{table}

\begin{table}[h] \centering 
   
\small
\begin{tabularx}{\columnwidth}{l @{\extracolsep{-1pt}} c @{\extracolsep{-1pt}} c @{\extracolsep{-1pt}} c @{\extracolsep{-1pt}} c} 
\\[-1.8ex]\hline 
\hline \\[-1.8ex] 
 & F1 & F2 & F3 & F4 \\ 
\hline \\[-1.8ex] 
1 & cousins & girl scout & paparazzi & homeschool \\ 
2 & clan & scout troop & auto & singles \\ 
3 & family group & troop & creations & arizona \\ 
4 & spring nail & scout & crafts & church of \\ 
5 & s family & kindergarten & custom & houston \\ 
6 & s usborne & mrs & parts & areas \\ 
7 & trip & grade & rocks & surrounding areas \\ 
8 & family & 20 21 & loving memory & county \\ 
9 & nail party & 2020 2021 & used & and surrounding \\ 
10 & extended & scouts & in loving & baptist church \\ 
11 & nail bar & pack & memory of & surrounding \\ 
12 & family page & 4 h & found & barter \\ 
13 & usborne & preschool & cars & items \\ 
14 & birthday & estates & baptist church & yardsale \\ 
15 & spring into & girl & small business & uncensored \\ 
16 & books more & neighborhood & car & buy nothing \\ 
17 & usborne books & 4th & who like & church \\ 
18 & descendants & subdivision & friends who & garage sale \\ 
19 & descendants of & neighbors & in memory & volunteers \\ 
20 & zyia party & 3rd & services & clothes \\ 
21 & s spring & hoa & sale or & yard sale \\ 
22 & virtual pampered & 2nd & memory & online yard \\ 
23 & bash & book club & or trade & pokemon \\ 
24 & chef party & committee & buy sale & garage \\ 
25 & into & 2020 & for sale & yard \\ 
\hline \\[-1.8ex] 
\end{tabularx} 

\caption{Group name ngrams with the highest average factor scores for each factor. See text for methodology. Only ngrams that are observed in at least 500 groups are included.}
\label{tbl:ngrams}
\vspace{-18pt}
\end{table} 

In Table~\ref{tbl:ngrams}, the salient ngrams for F1 support our interpretation of this factor. Family-related group ngrams such as ``cousins,'' ``clans,'' ``family group,'' ``descendants'' explain the tight-knit, but also geographically non-local nature of this factor. We speculate that in areas with high F1, geographically separated family groups are over-indexed.

Surprisingly, we also observe many multi-level marketing (MLM) related ngrams such as ``usborne books,'' ``pampered chef,'' ``zyia,'' etc. MLM is an industry that relies on a nonsalaried workforce making direct sales to customers. These nonsalaried sellers (also called consultants, distributors, etc.) take commissions out of their sales and also try to grow the seller network by recruiting downline consultants and taking a cut of these downline sales. MLM activity is known to rely on existing social capital and consultants need to tap a variety of bonding and bridging connections in their efforts to sell the products~\cite{lofthouse2021institutions}. 

While a closer analysis of the MLM phenomenon on Facebook groups is outside of the scope of this paper, the relation between MLMs and F1 is so strong that we were concerned that F1 could simply  be capturing MLM activity and related groups, without further ecological validity. As a robustness check, we went back and repeated our analyses from end to end, excluding groups that contain popular MLM brands and related keywords in their names (See Appendix for a full list of these keywords). The factor analysis results proved to be stable and we were still able to extract a similar factor that is over-indexed in the Midwest (see Appendix).

These results, taken together with the robustness check, are consistent with the idea that MLM activity relies on  the existing social capital stock of communities. In communities with higher trust and stronger bonds, people might find it easier to market the products and recruit downline consultants. The non-local nature of F1 is also consistent with MLM activity. We suspect that Facebook groups might be helping MLM participants to tap into their geographically long-distance ties to avoid hyper-local competition.

\subsubsection{F2: Very local and small groups}
The indicators that best measure F2 are the ratio of users who are in (very) local and (very) small groups (Figure ~\ref{fig:factor_loadings}). Fig.~\ref{fig:factor_cormat} further reveals F2 is not strongly correlated with population ($\rho=-0.04$), nor with density ($\rho=-0.19$) or rural population ($\rho=0$), and is very weakly correlated with community health ($\rho=0.23$) and its population-adjusted version ($\rho=0.21$). We do not observe any substantial correlation between F2 and any of the demographic or economic benchmark indicators ($\rho<0.30$ for all), but F2 is slightly negatively correlated county ethnic diversity ($\rho=-0.28$). 

Geographical pockets of the U.S. exhibit particularly low values of F2. New Mexico and Colorado, most of the South (Mississippi, Georgia, North and South Carolina, Virginia), Minnesota, and the San Francisco Bay Area in California, have low values of F2 (Figure \ref{fig:factor_choropleths}). 

Table~\ref{tbl:ngrams} shows a clear pattern of ngrams related to very local, neighborhood- and family-oriented groups: ``girl scout'' troops, ``neighborhood watch,'' ``HOA,'' (shorthand for Homeowners' Associations) ``kindergarten,'' ``preschool,'' ``soccer,'' ``mr'' and ``mrs'' (we suspect these titles are related to teachers of classes), etc.

It is somewhat surprising that a factor that seems to capture the prevalence of very local, neighborly group participation seems to exhibit virtually no correlation with social capital measures and a very weak correlation with external estimates of social capital and community health.

\subsubsection{F3: Public, age-diverse groups with 100+ members}

\begin{figure}[t]
\parbox{\columnwidth}{
\centering
\begin{subfigure}{1\columnwidth}
\includegraphics[width=\linewidth]{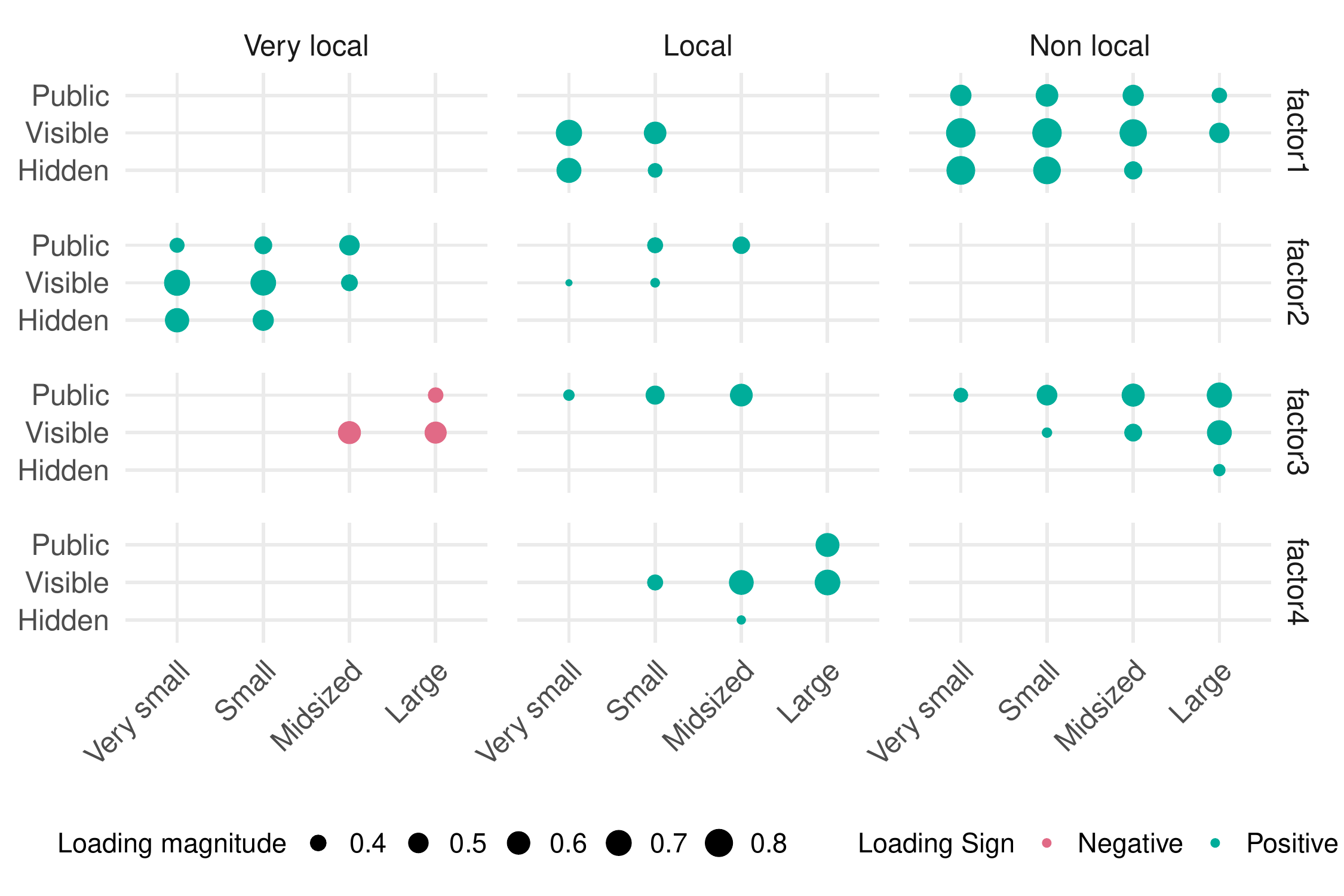}
\vspace{-12pt}
\subcaption{Group participation.}
\label{fig:factor_loadings_a}
\end{subfigure}
\begin{subfigure}{1\columnwidth}
\includegraphics[width=\linewidth]{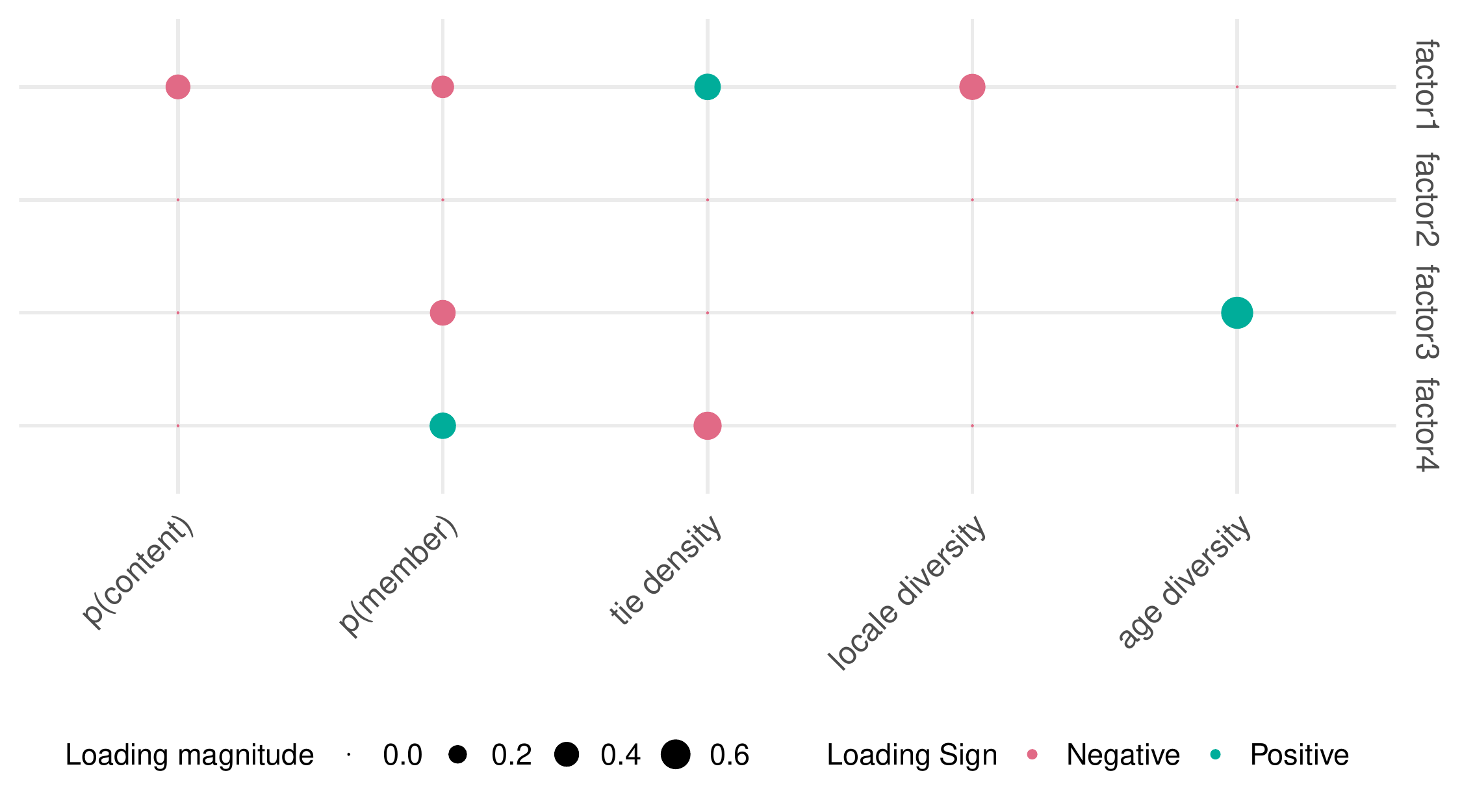}
\vspace{-12pt}
\subcaption{Local group characteristics.}
\label{fig:factor_loadings_b}
\end{subfigure}
\caption{Visual representation of indicator loadings on factors}
\label{fig:factor_loadings}
\vspace{-18pt}
}
\end{figure}

Indicators that contribute to F3 include age diversity (as captured by the average inter-quartile range of the members' age distribution in local groups) and the share of large and public groups and memberships (Fig.~\ref{fig:factor_loadings}). 

One could interpret the openness of groups, inter-age mixing, and participation in large groups, with access to many other individuals, as positively correlated with bridging social capital. That said, F3 is only weakly correlated with population-unadjusted community health (0.22). F3 also correlates positively with sparsely populated rural areas ($\rho=0.67$), which tend to have higher community health, and negatively with population and density. In fact, the correlation F3 has with community health changes signs once we adjust for population ($\rho=-0.19$). Thus, we believe F3 captures the contrast between the Facebook Groups usage patterns in small towns vs. urban centers. Population and rural population seem to be the mediator that drives the relation between this factor and community health. F3 is also strongly correlated with religious organizations per capita ($\rho=0.70$) and percentage of adults without a bachelor's degree ($\rho=-0.61$). 

We speculate that small towns may allow everyone to participate in fairly big, well-mixed settings. On the other hand, urban areas, where large, inclusive groups may grow too large for many purposes, could support smaller, more exclusive, and age-differentiated groups. 

The geographical distribution of F3 (Figure \ref{fig:factor_choropleths}) strengthens our interpretation: the value of this factor is elevated in parts of the Great Plains, Appalachia, and is depressed on the West Coast, North East Coast, and northern parts of the Midwest, including large urban centers.

A small puzzle is the contrasting result of positive loading of the percentage of people in private non-local and large groups ($0.701$) and negative loading of the percentage of users who are in private, very local, and large groups ($0.592$). Large and very local groups, by definition, require many members who live in the same county and this is possible in areas with a high population density. Large groups in rural areas, by necessity, span multiple counties and end up with low locality scores. This is a good reminder of the potential effects of our operationalization of locality which may have different meanings in rural and urban areas.

The n-gram analysis in Table~\ref{tbl:ngrams}, reveals a patchy pattern: ngrams related to transactional groups (buy and sell, services, used, etc.), ngrams that might indicate memorial groups (loving memory, in memory), and again some MLM-related keywords (``paparazzi'', an MLM brand, crafts). 

\subsubsection{F4: Partially local, member-controlled and bridging groups}

All group participation indicators (i.e., all but tie density and member gating) load on this factor with a positive weight (Figure ~\ref{fig:factor_loadings}). The indicators span different privacy and size buckets, but all of them are limited to local group types (but not non-local or very local groups). Especially midsized, large and local group participation indicators load on F4. In addition, this factor captures counties where groups have low friendship density among the members (loading -0.584) and do not involve membership gating procedures (loading 0.505). These findings suggest the factor is capturing partially local large group activity that brings people together who are otherwise not connected via friendship ties.

F4 has a slight correlation with population ($\rho=-0.33$) and a weak negative correlation with percentage of rural population ($\rho=-0.18$). The weak negative correlation between F4 and community health ($\rho=-0.23$) disappears when controlling for population ($\rho=0.08$). We don't observe any notable correlation with any of the other social benchmark indicators -- which, given this factor might be capturing local group activity, is notable on its own.

When we look at the most salient ngrams associated with F4, we see terms related to local associations (homeschool, church, volunteers) and terms that indicate a locality at large scales (surrounding areas) along with terms related to transactional groups (buy and sell, yardsale) and neighborhood groups that foster local gift economies (buy nothing).

Multiple regions score particularly low on F4: 1) Midwest and Great Plains, 2) Southern counties, 3) 1) parts of California, except most of Central Sierra Nevada, 4) parts of Virginia and North Carolina (Figure \ref{fig:factor_choropleths}). Low F4 scores in these regions indicate relative lack of memberships in local groups with sparse friendship ties that cover multiple counties.

\section{Discussion}
Our results reveal per-county geographic variation within the United States in participation in Facebook groups of various sizes, visibility, locality, age mixing, friendship density, and membership and content gating. We show that a substantial portion of the variability across all these dimensions can be represented by four main dimensions.

 F1, corresponding to small, private and non-localized groups, is one of the two factors correlated with offline social capital. It is highest in the Midwest, and its choropleth is visually similar to that of the JEC social capital index. This correspondence could be due to higher levels of social capital allowing for the formation of such groups, as friends can invite one another and sustain activity in the group without it necessarily needing to be local or open to all. An example of small groups relying on non-local friendship ties is that of multi-level marketing, though the results hold with such groups excluded. Furthermore, since counties with higher F1 have more non-religious organizations per capita, some group activity likely corresponds to participation in those formal organizations. F1 is also negatively correlated with the poverty rate.
 
 F2, representing participation in small or local Facebook groups, was not strongly correlated with offline indicators. This lack of association was also consistent with our initial exploration of participation in any local groups. This was a surprising result, if one expects areas with high social capital to have more interactions between neighbors. However, it may be that very local groups do not draw particularly on social capital, but rather depend on associations that may occur almost by default, e.g. families whose kids play on the same soccer team or are in the scouts, residents living in a neighborhood or building with a homeowners' association, etc.
 
 F3 captures interactions in more sparsely populated areas, where separation by age, or limiting participation by making a group private, makes less sense. F3 is negatively correlated with county population, median income, and \% of adults with a college degree, while being positively correlated with \% rural, and the number of religious organizations per capita. The positive association between F3 and the offline social capital indices reverses direction when we control for population and density of the counties.
 
 F4 highlights areas where people who are not already friends interact in partly local groups. That F4 is not correlated with offline social capital indices is again somewhat surprising, given that some of the groups it captures correspond to community organizations and activities such as volunteering, church, and homeschooling. Then again, it also captures transactional groups, such as for-sale groups, where social capital may not play as much of a role.
 
Although we have demonstrated strong correlation of online factors and informative grouping with respect to offline social capital indicators, we note that our analysis is based on differences in collective behavior on one platform, Facebook, on one product, Groups. Thus, it may contain biases due to differential adoption of Facebook and participation in groups by age and region.  Alternative online platforms that people use as a substitute to Facebook Groups may introduce similar biases. In this sense, we have limited data, but by providing the dataset to the research community we hope that future research will incorporate other sources and will be able see a more complete picture than we could at present.

\section{Conclusion and Future Work}
 In this paper we presented the first analysis of the correspondence between online group-based interaction in a given location, and social capital measures of the same.  The results are correlational. Most likely the offline environment is what we see reflected in online groups. On the other hand, it is possible that some communities are able to use social media, including Facebook, to foster greater social capital. For example, perhaps social media is able to support the kind of private, small groups that are more typical of rural areas, but for people living in dense areas. Similarly, finding new shared interests can overcome the lower trust that diverse communities have~\cite{putnam07diversity}. The incremental value that social media provides to social capital could be the subject of future work. 
 
 Another area of future exploration could utilize online indicators to detect changes in community social capital by region. More careful calibration with offline data, incorporating differential adoption, would be needed before one could potentially use Facebook variables to estimate changes in offline social capital. However, such indices could indeed be produced continuously, and only periodically calibrated with more expensive and time intensive surveys and offline data collection.

Given the promising results in this first analysis, we are sharing the Facebook group interaction dataset in the hopes that it may be incorporated in other studies of county-level differences, whether in measuring social capital or other inquiries, such as online collective action, civic engagement, or disaster response.

\balance{}

\bibliography{references}

\begin{thebibliography}{30}
\providecommand{\natexlab}[1]{#1}
\providecommand{\url}[1]{\texttt{#1}}
\providecommand{\urlprefix}{URL }
\expandafter\ifx\csname urlstyle\endcsname\relax
  \providecommand{\doi}[1]{doi:\discretionary{}{}{}#1}\else
  \providecommand{\doi}{doi:\discretionary{}{}{}\begingroup
  \urlstyle{rm}\Url}\fi

\bibitem[{Bailey et~al.(2018)Bailey, Cao, Kuchler, Stroebel, and
  Wong}]{bailey2018social}
Bailey, M.; Cao, R.; Kuchler, T.; Stroebel, J.; and Wong, A. 2018.
\newblock Social connectedness: Measurement, determinants, and effects.
\newblock \emph{Journal of Economic Perspectives} 32(3): 259--80.

\bibitem[{Bettencourt et~al.(2007)Bettencourt, Lobo, Helbing, K{\"u}hnert, and
  West}]{bettencourt2007growth}
Bettencourt, L.~M.; Lobo, J.; Helbing, D.; K{\"u}hnert, C.; and West, G.~B.
  2007.
\newblock Growth, innovation, scaling, and the pace of life in cities.
\newblock \emph{PNAS} 104(17): 7301--7306.

\bibitem[{Bj{\o}rnskov(2007)}]{bjornskov2007determinants}
Bj{\o}rnskov, C. 2007.
\newblock Determinants of generalized trust: A cross-country comparison.
\newblock \emph{Public choice} 130(1): 1--21.

\bibitem[{Burke, Kraut, and Marlow(2011)}]{burke2011social}
Burke, M.; Kraut, R.; and Marlow, C. 2011.
\newblock Social capital on Facebook: Differentiating uses and users.
\newblock In \emph{CHI'11}, 571--580.

\bibitem[{Ellison, Steinfield, and Lampe(2006)}]{ellison2006spatially}
Ellison, N.; Steinfield, C.; and Lampe, C. 2006.
\newblock Spatially bounded online social networks and social capital.
\newblock \emph{International Communication Association} 36(1-37).

\bibitem[{Ellison, Steinfield, and Lampe(2007)}]{ellison2007benefits}
Ellison, N.~B.; Steinfield, C.; and Lampe, C. 2007.
\newblock The benefits of Facebook “friends:” Social capital and college
  students’ use of online social network sites.
\newblock \emph{Journal of computer-mediated communication} 12(4): 1143--1168.

\bibitem[{Gastner and Newman(2006)}]{gastner2006optimal}
Gastner, M.~T.; and Newman, M.~E. 2006.
\newblock Optimal design of spatial distribution networks.
\newblock \emph{Physical Review E} 74(1): 016117.

\bibitem[{Gelman(2008)}]{gelman2008scaling}
Gelman, A. 2008.
\newblock Scaling regression inputs by dividing by two standard deviations.
\newblock \emph{Statistics in medicine} 27(15): 2865--2873.

\bibitem[{Granovetter(1985)}]{granovetter1985economic}
Granovetter, M. 1985.
\newblock Economic action and social structure: The problem of embeddedness.
\newblock \emph{American journal of sociology} 91(3): 481--510.

\bibitem[{Herda\u{g}delen, Adamic, and State(2022)}]{herdagdelen2023}
Herda\u{g}delen, A.; Adamic, L.; and State, B. 2022.
\newblock {Replication Data for: The Geography of Facebook Groups in the United
  States}.
\newblock \doi{10.7910/DVN/OYQVEP}.
\newblock \urlprefix\url{https://doi.org/10.7910/DVN/OYQVEP}.

\bibitem[{Iyer et~al.(2020)Iyer, Cheng, Brown, and Wang}]{iyer2020does}
Iyer, S.; Cheng, J.; Brown, N.; and Wang, X. 2020.
\newblock When Does Trust in Online Social Groups Grow?
\newblock In \emph{ICWSM}, volume~14, 283--293.

\bibitem[{La~Macchia et~al.(2016)La~Macchia, Louis, Hornsey, and
  Leonardelli}]{la2016small}
La~Macchia, S.~T.; Louis, W.~R.; Hornsey, M.~J.; and Leonardelli, G.~J. 2016.
\newblock In small we trust: Lay theories about small and large groups.
\newblock \emph{Personality and Social Psychology Bulletin} 42(10): 1321--1334.

\bibitem[{Lochner, Kawachi, and Kennedy(1999)}]{lochner1999social}
Lochner, K.; Kawachi, I.; and Kennedy, B.~P. 1999.
\newblock Social capital: a guide to its measurement.
\newblock \emph{Health \& place} 5(4): 259--270.

\bibitem[{Lofthouse and Storr(2021)}]{lofthouse2021institutions}
Lofthouse, J.~K.; and Storr, V.~H. 2021.
\newblock Institutions, the social capital structure, and multilevel marketing
  companies.
\newblock \emph{Journal of Institutional Economics} 17(1): 53--70.

\bibitem[{Ma et~al.(2019)Ma, Cheng, Iyer, and Naaman}]{ma2019people}
Ma, X.; Cheng, J.; Iyer, S.; and Naaman, M. 2019.
\newblock When Do People Trust Their Social Groups?
\newblock In \emph{CHI'19}, 1--12.

\bibitem[{Minkov and Hofstede(2012)}]{minkov2012hofstede}
Minkov, M.; and Hofstede, G. 2012.
\newblock Hofstede’s fifth dimension: New evidence from the World Values
  Survey.
\newblock \emph{Journal of cross-cultural psychology} 43(1): 3--14.

\bibitem[{Neville(2012)}]{Neville_2012}
Neville, L. 2012.
\newblock Do economic equality and generalized trust inhibit academic
  dishonesty? Evidence from state-level search-engine queries.
\newblock \emph{Psychological Science} 23(4): 339–345.

\bibitem[{Newton(2001)}]{newton2001trust}
Newton, K. 2001.
\newblock Trust, social capital, civil society, and democracy.
\newblock \emph{International political science review} 22(2): 201--214.

\bibitem[{Putnam(2007)}]{putnam07diversity}
Putnam, R.~D. 2007.
\newblock E Pluribus Unum: Diversity and Community in the 21st Century.
\newblock \emph{Scandinavian Political Studies} 30(2): 137--174.

\bibitem[{Putnam et~al.(2000)}]{putnam2000bowling}
Putnam, R.~D.; et~al. 2000.
\newblock \emph{Bowling alone: The collapse and revival of American community}.
\newblock Simon and schuster.

\bibitem[{Scott and Johnson(2005)}]{scott2005bowling}
Scott, J.~K.; and Johnson, T.~G. 2005.
\newblock Bowling alone but online together: Social capital in e-communities.
\newblock \emph{Community Development} 36(1): 9--27.

\bibitem[{Simpson(2006)}]{simpson2006poverty}
Simpson, B. 2006.
\newblock The poverty of trust in the Southern United States.
\newblock \emph{Social Forces} 84(3): 1625--1638.

\bibitem[{{Social Capital Project}(2018)}]{JEC2018}
{Social Capital Project}. 2018.
\newblock {The Geography of Social Capital in America}.
\newblock Technical Report 1-18, Joint Economic Committee - Republicans.
\newblock
  \urlprefix\url{https://www.jec.senate.gov/public/index.cfm/republicans/socialcapitalproject}.

\bibitem[{Steinfield, Ellison, and Lampe(2008)}]{steinfield_social_2008}
Steinfield, C.; Ellison, N.~B.; and Lampe, C. 2008.
\newblock Social capital, self-esteem, and use of online social network sites:
  {A} longitudinal analysis.
\newblock \emph{Journal of Applied Developmental Psychology} 29(6): 434--445.

\bibitem[{Stolle(2002)}]{stolle2002trusting}
Stolle, D. 2002.
\newblock Trusting strangers--the concept of generalized trust in perspective.
\newblock \emph{{\"O}sterreichische Zeitschrift f{\"u}r Politikwissenschaft}
  31(4): 397--412.

\bibitem[{Tiwari, Lane, and Alam(2019)}]{tiwari2019social}
Tiwari, S.; Lane, M.; and Alam, K. 2019.
\newblock Do social networking sites build and maintain social capital online
  in rural communities?
\newblock \emph{Journal of rural studies} 66: 1--10.

\bibitem[{Vanden~Abeele(2018)}]{vanden2018does}
Vanden~Abeele, e.~a. 2018.
\newblock Does Facebook use predict college students’ social capital?
\newblock \emph{Communication Studies} 69(3): 272--282.

\bibitem[{Wellman, Boase, and Chen(2002)}]{wellman2002networked}
Wellman, B.; Boase, J.; and Chen, W. 2002.
\newblock The networked nature of community: Online and offline.
\newblock \emph{It \& Society} 1(1): 151--165.

\bibitem[{Williamson(2008)}]{williamson2008transaction}
Williamson, O.~E. 2008.
\newblock Transaction cost economics.
\newblock In \emph{Handbook of new institutional economics}, 41--65. Springer.

\bibitem[{Wu(2020)}]{wu2020does}
Wu, C. 2020.
\newblock Does migration affect trust? Internal migration and the stability of
  trust among americans.
\newblock \emph{The Sociological Quarterly} 61(3): 523--543.

\end{thebibliography}
\newpage
\appendix
\section{Appendix}

\subsection{Group Size Distribution}

\begin{figure}[H]
\centering
\includegraphics[width=\columnwidth]{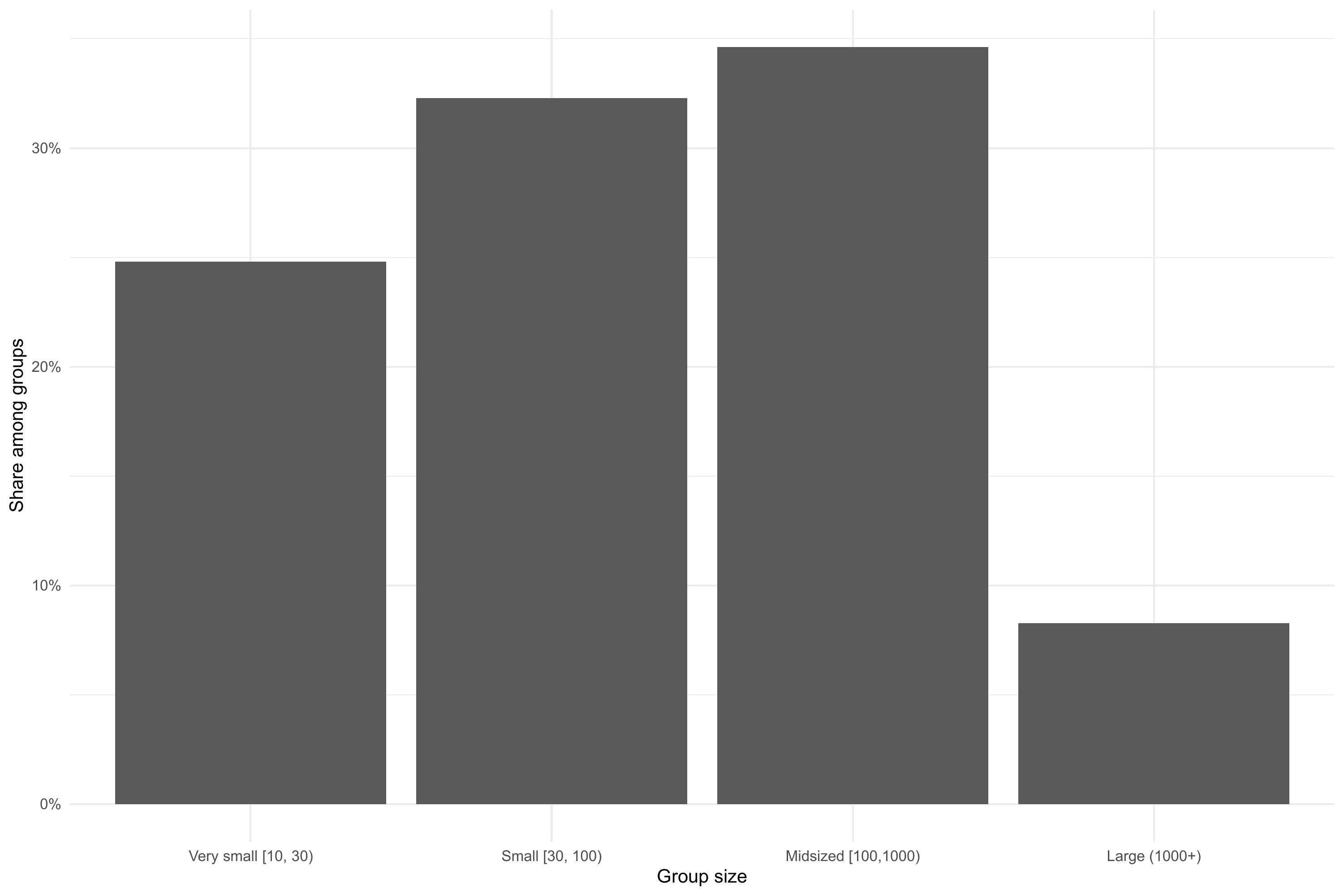}
\caption{Share of group size buckets}
\label{fig:group_size}
\end{figure}

As seen in Fig.~\ref{fig:group_size}, the threshold we picked for size buckets give roughly equal-sized buckets in terms of groups with a smaller bucket of large groups, capturing the tail-end of the distribution.

\subsection{County-level Aggregate Indicators}
Group size, locality, and privacy together specify 36 mutually exclusive and exhaustive combinations a given Facebook group can be placed in (4 size buckets * 3 locality buckets * 3 privacy types). Each combination leads to a separate group participation indicator that is aggregated at the county level. In addition to these 36 indicators, we also include 6 aggregate indicators that characterize the local groups observed in each county (age diversity, gender diversity, linguistic diversity, tie density, and content and membership controls). Taken together, we end up with 42 indicators at the county level that summarize the group participation characteristics in a given county.

In Table~\ref{tbl:descriptives}, we provide average proportion of Monthly Active Users, for various group characteristics.

\setlength{\tabcolsep}{3pt}
\begin{table}[H]
\centering
\small
\begin{tabular}{| l | l | l | d{2} | d{2} | d{2} | d{2} | d{2} |}
  \hline
  & Size & Privacy & \multicolumn{1}{c |}{Mean} & \multicolumn{1}{c |}{S.D.} & \multicolumn{1}{c |}{Med.} & \multicolumn{1}{c |}{Min} & \multicolumn{1}{c |}{Max} \\ 
  \hline
\multirow{13}{1em}{\rotatebox[origin=c]{90}{Non-local}} & \multirow{3}{2em}{Very small} & Public & <.01 & <.01 & <.01 & .00 & .01 \\ 
   &  & Visible & .02 & <.01 & .01 & <.01 & .07 \\ 
   &  & Hidden & <.01 & <.01 & <.01 & <.01 & .03 \\ 
\cline{2-8}
   & \multirow{3}{2em}{Small} & Public & .01 & <.01 & .01 & <.01 & .06 \\ 
   &  & Visible & .04 & .01 & .04 & <.01 & .11 \\ 
   &  & Hidden & .02 & <.01 & .02 & <.01 & .05 \\ 
\cline{2-8}
   & \multirow{3}{2em}{Mid-size} & Public & .07 & .02 & .06 & <.01 & .40 \\ 
   &  & Visible & .12 & .03 & .12 & .02 & .32 \\ 
   &  & Hidden & .02 & <.01 & .02 & <.01 & .15 \\ 
\cline{2-8}
   & \multirow{3}{2em}{Large} & Public & .20 & .06 & .19 & .03 & .47 \\ 
   &  & Visible & .29 & .06 & .29 & .08 & .60 \\ 
   &  & Hidden & .04 & .02 & .04 & <.01 & .19 \\ 
\cline{2-8}
 & \multicolumn{2}{c}{Overall} & .83 & .83 & .19 & .17 & 1.62 \\ \hline
  \multirow{13}{1em}{\rotatebox[origin=c]{90}{Local}} & \multirow{3}{2em}{Very small} & Public & <.01 & <.01 & <.01 & .00 & .02 \\ 
   &  & Visible & .02 & <.01 & .01 & <.01 & .06 \\ 
   &  & Hidden & <.01 & <.01 & <.01 & <.01 & .05 \\ 
\cline{2-8}
   & \multirow{3}{2em}{Small} & Public & .01 & <.01 & <.01 & <.01 & .07 \\ 
   &  & Visible & .04 & .01 & .04 & <.01 & .15 \\ 
   &  & Hidden & .01 & <.01 & .01 & .00 & .05 \\ 
\cline{2-8}
   & \multirow{3}{2em}{Mid-size} & Public & .07 & .03 & .06 & <.01 & .51 \\ 
   &  & Visible & .09 & .03 & .09 & .02 & .29 \\ 
   &  & Hidden & .01 & <.01 & .01 & .00 & .20 \\ 
\cline{2-8}
   & \multirow{3}{2em}{Large} & Public & .12 & .06 & .12 & .01 & .43 \\ 
   &  & Visible & .16 & .07 & .15 & .02 & .45 \\ 
   &  & Hidden & .01 & .02 & <.01 & .00 & .24 \\ 
\cline{2-8}
 & \multicolumn{2}{c}{Overall} &   .54 & .55 & .17 & .10 & 1.27 \\ \hline
  \multirow{13}{1em}{\rotatebox[origin=c]{90}{Very local}} & \multirow{3}{2em}{Very small} & Public & <.01 & <.01 & <.01 & .00 & .01 \\ 
   &  & Visible & <.01 & <.01 & <.01 & .00 & .04 \\ 
   &  & Hidden & <.01 & <.01 & <.01 & .00 & .04 \\ 
\cline{2-8}
   & \multirow{3}{2em}{Small} & Public & <.01 & <.01 & <.01 & .00 & .03 \\ 
   &  & Visible & <.01 & <.01 & <.01 & .00 & .09 \\ 
   &  & Hidden & <.01 & <.01 & <.01 & .00 & .04 \\ 
\cline{2-8}
   & \multirow{3}{2em}{Mid-size} & Public & <.01 & <.01 & <.01 & .00 & .08 \\ 
   &  & Visible & .01 & .01 & <.01 & .00 & .12 \\ 
   &  & Hidden & <.01 & <.01 & <.01 & .00 & .06 \\ 
\cline{2-8}
   & \multirow{3}{2em}{Large} & Public & <.01 & <.01 & <.01 & .00 & .11 \\ 
   &  & Visible & <.01 & .01 & <.01 & .00 & .12 \\ 
   &  & Hidden & <.01 & <.01 & <.01 & .00 & .06 \\ 
\cline{2-8}
 & \multicolumn{2}{c}{Overall} &   .05 & .04 & .04 & <.01 & .37 \\ 
   \hline
\end{tabular}
\caption{County-level aggregated indicators. Statistics over average proportion of Monthly Active Users in each county, for each type of groups with specific characteristics.}
\label{tbl:descriptives}
\end{table}

\subsection{Indicator Correlation Matrix}
\begin{figure*}[hp]
\centering
 \includegraphics[width=\linewidth]{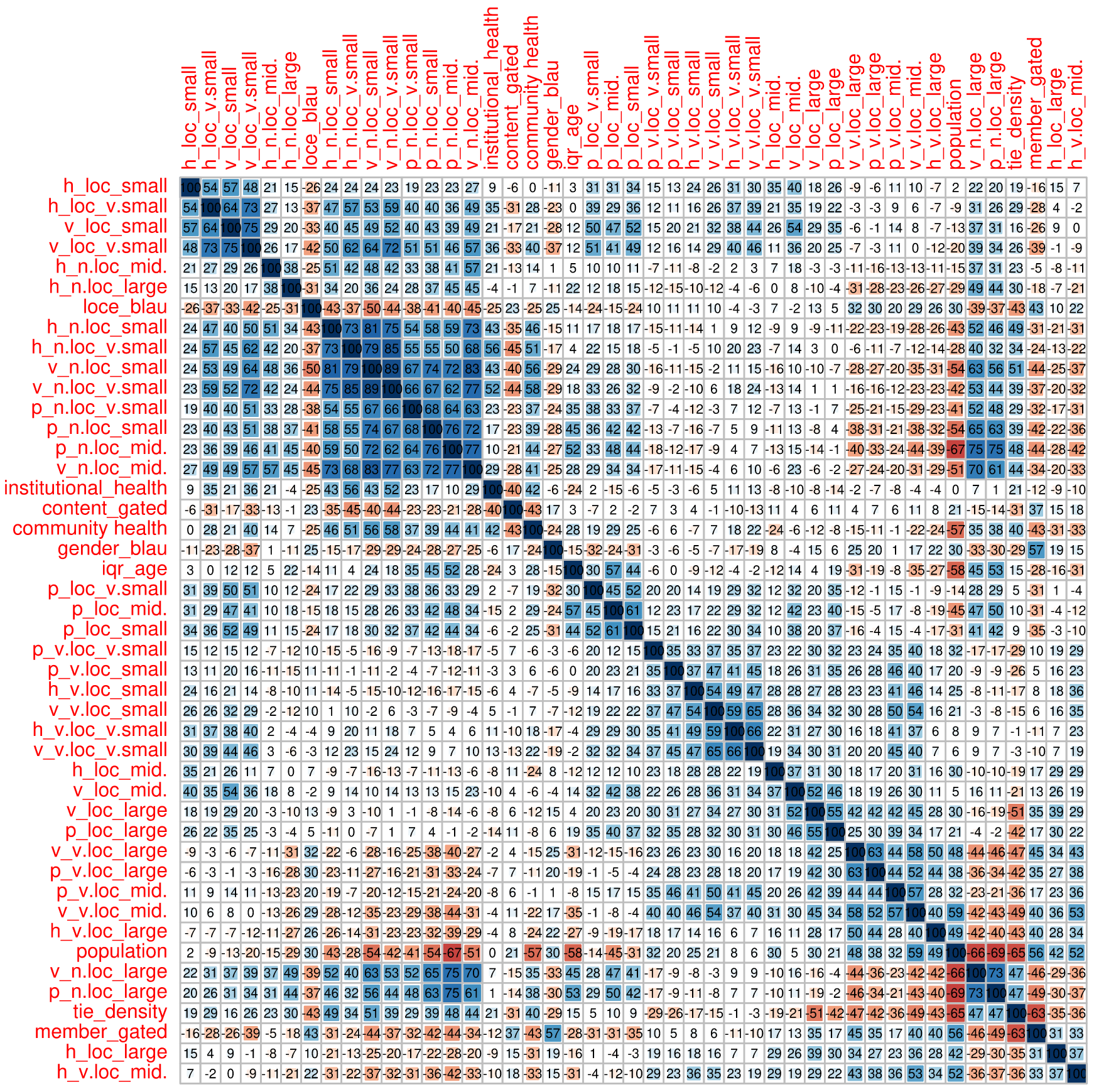}
\caption{Spearman correlation table of indicators and the unadjusted and adjusted community health, ordered on the absolute values by the hclust method in the corrplot:corrMatOrder function in R. Prefixes p, v, h correspond to public, visible, hidden. loc, v.loc, and n.loc correspond to local, very local and non-local.
}
\label{fig:corrmatrix}
\end{figure*}

The full correlation matrix (Figure~\ref{fig:corrmatrix}) of the 42 group indicators (and the two external indicators, community health and its population-adjusted version) reveals a multi-factorial internal structure. We can visually identify groups of indicators that are strongly (either positively or negatively) correlated with each other. In the main analysis, we follow a more rigorous approach to identify the latent factors that can explain these correlations.

\newpage
\subsection{Factor analysis technical details}

Parallel analysis and optimal coordinates heuristics suggested to use four factors. We tried 3-factor and 5-factor solutions and although the results were not qualitatively very different, we find the 4-factor solution more interpretable.

\begin{figure}[htb]
\centering
\includegraphics[width=0.95\linewidth]{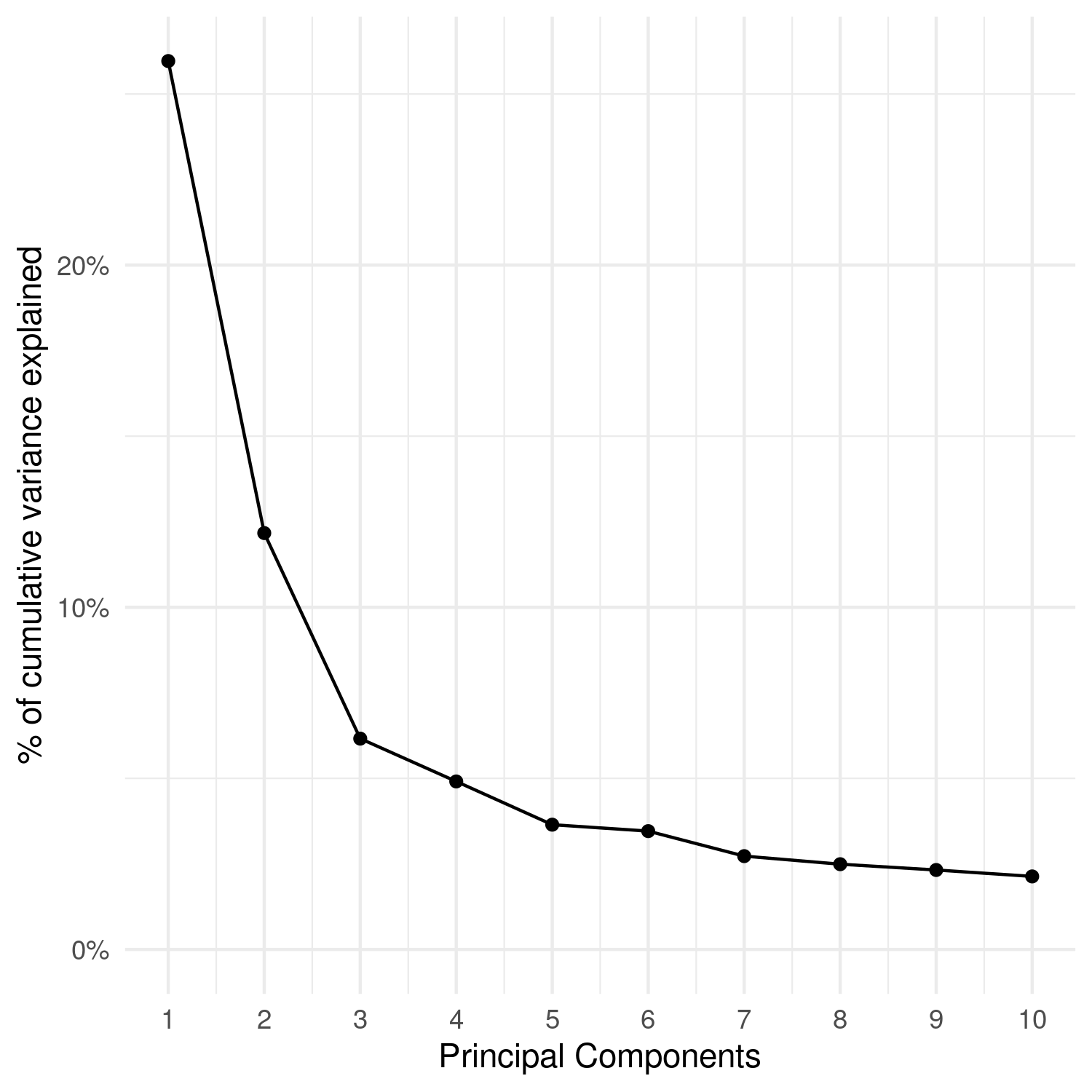}
\caption{Percentage of variance explained by principal components. We used a four-factor solution, suggested by parallel analysis (using the nFactors::nScree function in R)}
\label{fig:scree}
\end{figure}

\subsection{Factor loadings}
In Tables~\ref{tbl:f1_loadings}, \ref{tbl:F2_loadings}, \ref{tbl:F3_loadings}, and \ref{tbl:f4_loadings}, we provide the factor loadings of the indicators. A cut-off of absolute 0.30 loading is used for the loading tables.

\begin{table}[H] \centering 
  \caption{Online indicators loading on F1.} 
  \label{tbl:f1_loadings} 

\begin{tabular}{@{\extracolsep{5pt}} cc} 
\\[-1.8ex]\hline 
\hline \\[-1.8ex] 
visible\_non\_local\_very\_small & $0.898$ \\ 
visible\_non\_local\_small & $0.895$ \\ 
hidden\_non\_local\_very\_small & $0.860$ \\ 
hidden\_non\_local\_small & $0.806$ \\ 
visible\_non\_local\_midsized & $0.791$ \\ 
visible\_local\_very\_small & $0.741$ \\ 
hidden\_local\_very\_small & $0.684$ \\ 
public\_non\_local\_small & $0.590$ \\ 
visible\_local\_small & $0.590$ \\ 
public\_non\_local\_very\_small & $0.546$ \\ 
public\_non\_local\_midsized & $0.542$ \\ 
visible\_non\_local\_large & $0.516$ \\ 
group\_avg\_tie\_density & $0.479$ \\ 
group\_avg\_locale\_blau & $$-$0.466$ \\ 
hidden\_non\_local\_midsized & $0.452$ \\ 
control\_p\_content\_gated & $$-$0.415$ \\ 
public\_non\_local\_large & $0.392$ \\ 
hidden\_local\_small & $0.378$ \\ 
control\_p\_member\_gated & $$-$0.330$ \\ 
\hline \\[-1.8ex] 
\end{tabular} 

\end{table}

\begin{table}[H] \centering 
  \caption{Online indicators loading on F2.} 
  \label{tbl:F2_loadings} 

\begin{tabular}{@{\extracolsep{5pt}} cc} 
\\[-1.8ex]\hline 
\hline \\[-1.8ex] 
visible\_very\_local\_very\_small & $0.768$ \\ 
visible\_very\_local\_small & $0.748$ \\ 
hidden\_very\_local\_very\_small & $0.689$ \\ 
hidden\_very\_local\_small & $0.568$ \\ 
public\_very\_local\_midsized & $0.544$ \\ 
public\_very\_local\_small & $0.469$ \\ 
public\_local\_midsized & $0.457$ \\ 
visible\_very\_local\_midsized & $0.444$ \\ 
public\_local\_small & $0.423$ \\ 
public\_very\_local\_very\_small & $0.404$ \\ 
visible\_local\_small & $0.335$ \\ 
visible\_local\_very\_small & $0.328$ \\ 
public\_very\_local\_large & $0.318$ \\ 
\hline \\[-1.8ex] 
\end{tabular} 

\end{table}

\begin{table}[H] \centering 
  \caption{Online indicators loading on F3.} 
  \label{tbl:F3_loadings} 

\begin{tabular}{@{\extracolsep{5pt}} cc} 
\\[-1.8ex]\hline 
\hline \\[-1.8ex] 
group\_avg\_iqr\_age & $0.760$ \\ 
public\_non\_local\_large & $0.720$ \\ 
visible\_non\_local\_large & $0.701$ \\ 
public\_non\_local\_midsized & $0.629$ \\ 
visible\_very\_local\_midsized & $$-$0.624$ \\ 
public\_local\_midsized & $0.616$ \\ 
visible\_very\_local\_large & $$-$0.592$ \\ 
public\_non\_local\_small & $0.542$ \\ 
public\_local\_small & $0.492$ \\ 
control\_p\_member\_gated & $$-$0.463$ \\ 
visible\_non\_local\_midsized & $0.458$ \\ 
public\_very\_local\_large & $$-$0.407$ \\ 
public\_non\_local\_very\_small & $0.390$ \\ 
hidden\_non\_local\_large & $0.353$ \\ 
public\_local\_very\_small & $0.343$ \\ 
visible\_non\_local\_small & $0.334$ \\ 
group\_avg\_tie\_density & $0.320$ \\ 
\hline \\[-1.8ex] 
\end{tabular} 

\end{table}

 \begin{table}[H] \centering 
   \caption{Online indicators loading on F4.} 
   \label{tbl:f4_loadings} 
 
\begin{tabular}{@{\extracolsep{5pt}} cc} 
\\[-1.8ex]\hline 
\hline \\[-1.8ex] 
visible\_local\_large & $0.757$ \\ 
visible\_local\_midsized & $0.717$ \\ 
public\_local\_large & $0.680$ \\ 
group\_avg\_tie\_density & $$-$0.584$ \\ 
control\_p\_member\_gated & $0.505$ \\ 
visible\_local\_small & $0.430$ \\ 
hidden\_local\_midsized & $0.337$ \\ 
hidden\_local\_large & $0.320$ \\ 
visible\_very\_local\_midsized & $0.307$ \\ 
public\_local\_midsized & $0.305$ \\ 
hidden\_local\_small & $0.301$ \\ 
\hline \\[-1.8ex] 
\end{tabular} 

 \end{table}
 
 \subsection{Conditional Effects on F1 Model}

\begin{figure}[H]
\centering
\includegraphics[width=\linewidth]{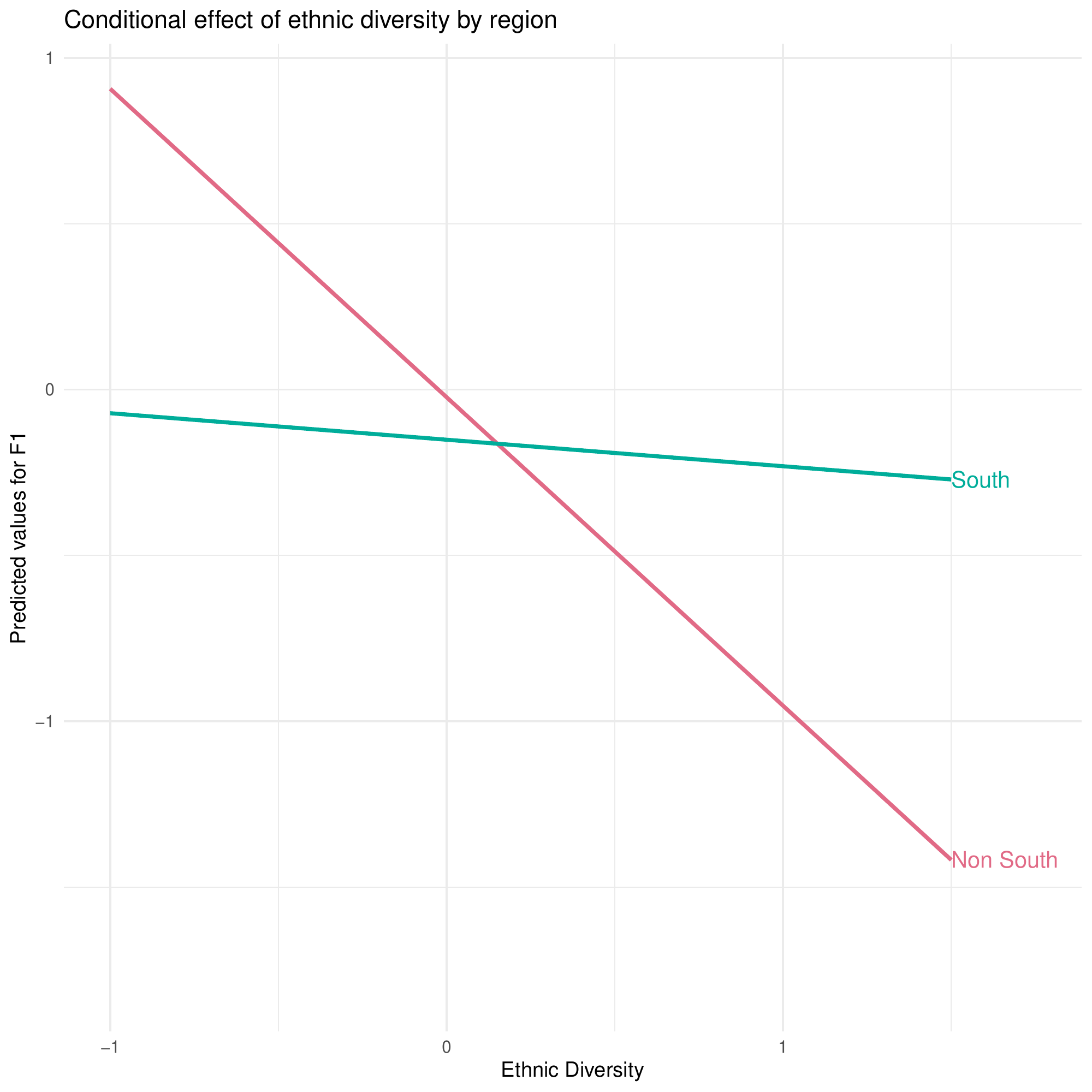}
\caption{Effect of standardized ethnic diversity conditioned on region in a linear model. Other control variables, population, religious organizations per capita, median household income, percentage of rural population and percentage of adults who graduated high school are fixed at their mean values.}
\label{fig:lm_f1_residuals}
\end{figure}

As mentioned in the main text, the divergence of the relation between ethnic diversity and F1 in Southern and other counties is visible in Figure~\ref{fig:lm_f1_residuals}. In this figure, we plot the residual of F1 in a linear model using only the control covariates (population, education, income, religion), as a function of ethnic diversity for Southern and other counties. 

\subsection{Multilevel Marketing Groups}
The presence of MLM-related keywords among the top group name ngrams associated with F1 prompted us to re-run our analyses end to end excluding groups whose names match common MLM keywords. This keyword list is not meant to be exhaustive, but we believe it captures a significant portion of MLM brands that are active on Facebook.

\begin{table}[H] \centering
\caption{Keywords used to exclude MLM-related groups based on group names}
\label{mlm_keywords}
\begin{tabular}{c c c}
arbonne&marykay&pinkzebra\\
beauty counter&melaleuca&pure romance\\
beautycounter&monat&rodan\\
color street&mythirtyone&scentsy\\
colorstreet&nail bar&senegence\\
crunchi&nail party&tastefully simple\\
dotdotsmile&neways&touchstone crystal\\
doterra&noonday&touchstonecrystal\\
epicure&norwex&tupperware\\
farmasi&optavia&unicity\\
flexship&origami owl&usborne\\
herbalife&origamiowl&xango\\
jamberry&pampered&younique\\
lemongrass&paparazzi&zija\\
lipsense&park lane&zyia\\
lularoe&parklane&\\
mary kay&pink zebra&\\	
\end{tabular}
\end{table}

\begin{figure}[H]
\centering
\includegraphics[width=\linewidth]{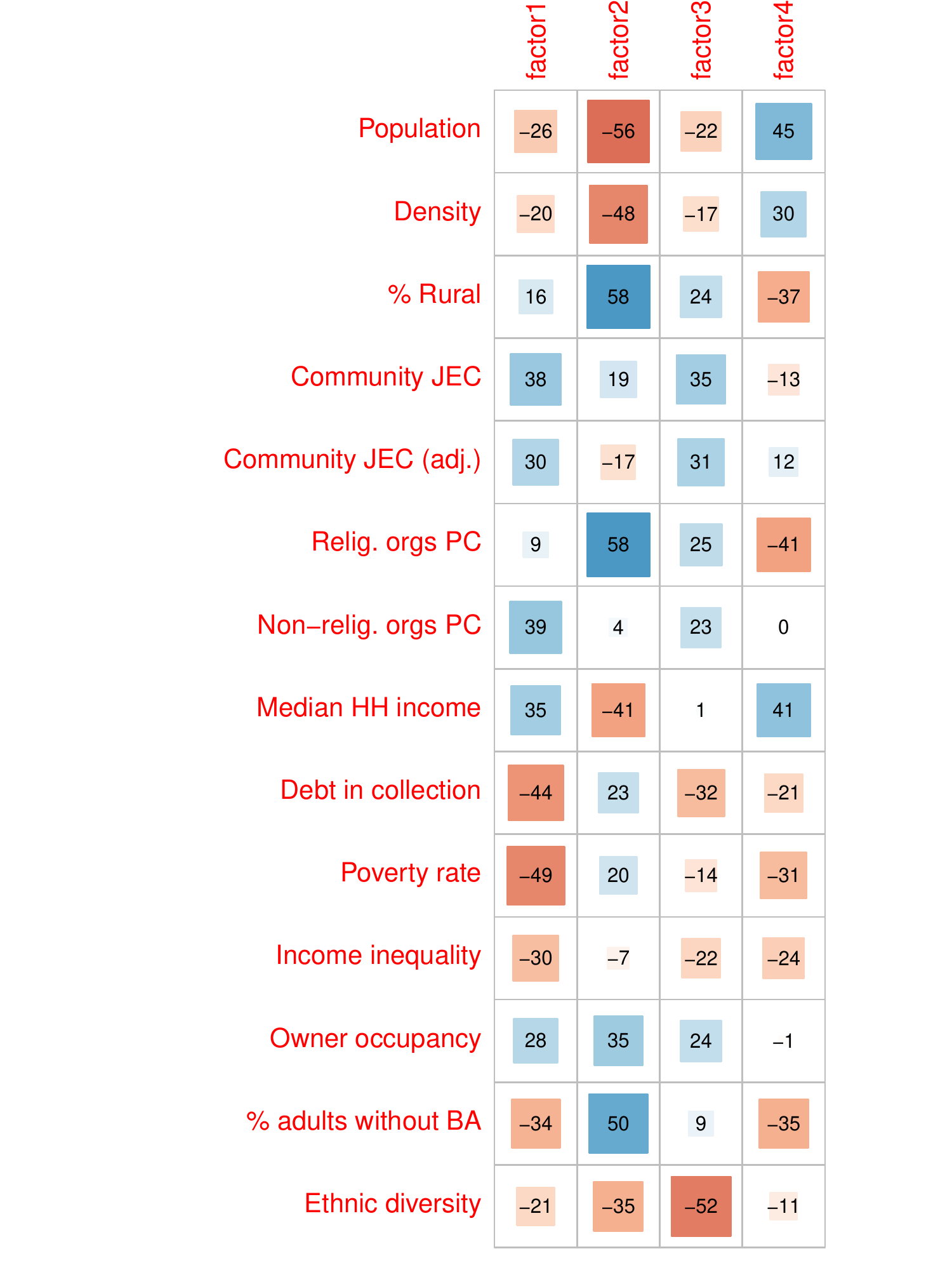}
\vspace{-12pt}
\caption{Spearman correlation between the factors and societal indicators, excluding MLM-related groups.}
\label{fig:factor_cormat_exmlm}
\vspace{-12pt}
\end{figure}

The factor analysis yields similar results. The correlations between the factors and benchmark indicators are given in Fig.\ref{fig:factor_cormat_exmlm}. Of interest to us, F1 in the MLM-excluding specification, has comparable correlation coefficients with both unadjusted ($\rho=0.38$) and population-adjusted community health ($\rho=0.30$) compared to F1 extracted in the main analysis. Among the two major components of JEC's measure, F1 (excluding MLM) is correlated with non-religious, non-profit organizations per capita ($\rho=0.39$), but not with religious organizations per capita ($\rho=0.09$), replicating the results we obtained in the main analysis.

\end{document}